\documentclass[pra,aps,showpacs,twocolumn,a4paper,tightenlines]{revtex4}
\usepackage{amsmath,amsfonts,amssymb,mathrsfs,bm}
\usepackage{verbatim,psfrag,times,CJK}
\usepackage{graphicx,graphics,color,epsfig}

\definecolor{red}{rgb}{1,0,0}
\usepackage{indentfirst}
\newcommand{\ket}[1]{| #1 \rangle}
\newcommand{\bra}[1]{\langle #1 |}

\begin{document}
\title{Preparation of a single-photon dark state in a chiral quantum system}

\author{Fengzheng Zhu, Teng Zhao, Hebin Zhang}
\author{Gao-xiang Li}
\email{gaox@phy.ccnu.edu.cn}
\affiliation{Department of Physics, Huazhong Normal University, Wuhan 430079, Peoplel's Republic of China}
\author{Zbigniew Ficek$^{a,b}$}
\affiliation{$^{a}$The National Center for Applied Physics, KACST, P.O. Box 6086, Riyadh 11442, Saudi Arabia\\
$^{b}$Quantum Optics and Engineering Division, Institute of Physics, University of Zielona G\'ora, Szafrana 4a, Zielona G\'ora 65-516, Poland}
%\author{$\ast$}
%\altaffiliation{$\ast$}

\begin{abstract}
We examine conditions under which an open quantum system composed of a driven degenerated parametric oscillator cavity and a driven two-level atom coupled to a waveguide could decay to a pure dark state rather than the expected mixed state. The calculations are carried out analytically in a low dimensional Hilbert space truncated at the double-excitation states of the combined system. The validity of the truncation is confirmed by the exact numerical analysis.
It is found that one way to produce the pure state is to chirally couple the cavity and the atom to the waveguide. Another way to produce the pure state is to drive the cavity and the atom with unequal detunings. In both cases, if the driving fields are weak, the produced state is a coherent superposition of only the single-excitation and ground states of the combined system. In addition, we have found a direct correspondence between the generation of the dark state and the photon blockade effect. In other words, the generation of the dark state acts as a blockade to the number of photons so that only a single photon can be present in the cavity. We investigate the normalized second-order correlation function of the cavity field and find that the conditions under which the correlation function vanishes coincide with the conditions for the creation of the pure dark state. This system is, therefore, suggested as an alternative scheme for the generation of single-photon states.
\end{abstract}

\date{\today}

\pacs{42.50.Ex, 03.67.Bg, 42.50.Ct, 42.79.Gn}
\maketitle

\section{introduction}\label{sec1}

Chirality in quantum optics refers to a situation in which two open quantum systems coupled to two counter-propagating modes of a waveguide interact in such a way that one system reacts to photons emitted by the other, while there is no interaction in the reverse direction~\cite{lm16,vr16}. Such directional interaction appears naturally in nanophotonic structures such as waveguides and optical nanofibers, where light experiences
tight transverse confinement. It can also be achieved by adding the spin-orbit coupling to the one-dimensional (1D) quasi-BEC or by controlling the spin orientation. This form of coupling has already been realized experimentally to control the direction of photon emission~\cite{pv14,ms14,sn15,yt15,nb14}.

The concept of chiral coupling between quantum systems has led to a new treatment of the radiation-matter interaction. For example, spontaneous emission from excited atoms and the radiative interaction between distant atoms may be considerably modified by the chiral coupling of the atoms to a photonic waveguide playing a role of a reservoir. In particular, the chiral coupling in a driven-dissipative multi-atom system can lead to the formation of dimers composed of two neighboring atoms which is manifested by the presence of a subradiant (dark) state of the system~\cite{rp14,pr15,sr12}. The dark state is completely decoupled from the reservoir and the external driving field which results in locking of the atomic spins. This represents a novel phenomenon of dissipative quantum magnetism and has been suggested as a new possibility to generate atom-atom entanglement in multiple qubit systems~\cite{ki14}. The formation of dimers by the chiral coupling has also been investigated in a system composed of a mixture of two-species of cold quantum gases, where one species appears as the emitter and the other one plays the role of one-dimensional phononic reservoir~\cite{rp14}. In addition, this concept has been applied to a system composed of a collection of two-level atoms and two Kerr nonlinear cavities coupled to a one-dimensional photonic waveguide~\cite{sr12}.
Apart from the basic interest in modifying the coupling between systems, the study of chirality is also motivated by potential applications to quantum information and the development of quantum computing~\cite{hk08}.

Another useful aspect of the chiral coupling is the control of spontaneous decay of quantum systems to a reservoir. In fact, the chiral-type coupling of quantum systems to a common reservoir is an analog of a cascaded quantum system, where the quantum systems are directionally coupled one to another without the backward coupling~\cite{wg93,hc93,kc94,gp94,nc05,gz04}.
In addition, interesting studies have been devoted to the phenomenon of photon blockade which yields to the creation of optical single-photon sources. Some nonlinearity, not necessarily strong, should be present in a given system leading to a blockade of the transmission of more than one photon through the system.
In recent years, the photon blockade effect has been the subject of numerous theoretical studies and various systems and mechanisms have been considered such as cavity quantum electrodynamics, quantum dots, and quantum optomechanical systems~\cite{ki11,ki12}. At the same time, experimental studies have been making progress in different systems as well, such as the optical cavities coupled to a trapped atom, circuit cavity QED, and quantum-dot in a photonic crystal~\cite{ki13}.

The notion of blockade of multi-photon excitations is not restricted to photon blockade for the transport of light through an optical system~\cite{is97,hs11,bb05,tc92,bi11,kk16,ki01,ki03,ki05,ki06}, but it has been extended to the dipole-dipole blockade of the simultaneous excitation of atoms~\cite{cp10,lf01,uj09,ga10,ge10,hh10,zm10,nm10,ss11,at11,jr10,gm09,hr07}, and to the Coulomb blockade of resonant transport of electrons through small metallic or semiconductor devices~\cite{al86,fd87,mk92}.
In analogy to photon blockade, the dipole-dipole blockade prevents absorption of more than one photon by a multi-atom system, and the Coulomb blockade prevents transport of more than one electron through a metallic or semiconductor device.

In this paper, we examined the dynamics of a system composed of a two-level atom and a single-mode degenerate parametric oscillator (DPO) cavity coupled to a waveguide. Specifically, we concentrate on determining conditions under which the system decays to a pure dark state. In particular, we describe how one could use a chiral quantum network to create the pure dark state in the system. We show that the stationary dark state, if it exists, belongs to a decoherence-free subsystem of the Hilbert space that is not affected by any environment-induced non unitary dynamics~\cite{ps96,dg97,zr97,lc98,kl00,pb07} and which under certain conditions does not evolve. We find two alternative ways to generate the pure state in the system. One way is to chirally couple the cavity and the atom to the waveguide. Another way is to drive the atom and the cavity mode with laser fields of unequal detunings. We approach the problem first using only the single-excitation basis and then extend the calculations to the double-excitation basis. In both cases we give an analytic description which accounts for the conditions for the system to collapse into a pure dark state. We find that in both cases, this particular state is a coherent superposition of the ground and single-photon excited states only.
We next show that there is a direct relationship between the generation of the single-photon dark state and the photon blockade effect. The ability to see the possibility to generate a single-photon state is enabled by calculating the normalized second-order correlation function which vanished under the conditions for the generation of the dark state. This effect is a particularly interesting feature of coupled systems as it demonstrates that the composite system can effectively behave as a two-level system.

The paper is organized as follows. In Sec.~\ref{sec2}, the master equation is introduced. We rewrite the master equation in terms of collective operators and determine a convenient basis for the calculations. Certain properties of the collective operators are discussed. We then distinguish spontaneous decay channels in the system and point out the existence of a single-excitation state which does not decay.
In Sec.~\ref{sec3}, we obtain the conditions for decoherence-free subspace and dark state in the Hilbert space of the system truncated at the single- and double-excitation states.
In Sec.~\ref{sec4}, we explore the connection between the dark state production and the photon blockade effect. We calculate numerically, without the truncation of the Hilbert space, the normalized second-order correlation function of the cavity field and show that at the dark state conditions the correlation function vanishes which is the clear indication that no more than one photon is present in the cavity field. This also demonstrates the validity of the truncation of the Hilbert space at the double-excitation states for a weak driving. Finally, in Sec.~\ref{sec5}, we summarize our results.
The paper concludes with three appendices in which certain aspects of the calculations are explained in more details. In Appendix~\ref{App1}, we give details of the derivation of the master equation for the reduced density operator of the system. The representation of the collective operators in the basis of the product states is given in Appendix~\ref{App2}. In Appendix~\ref{App3}, we give the complete set of equations of motion for the density matrix elements.

\section{The system and its evolution}~\label{sec2}	

\begin{figure}[h]
\centering
\includegraphics[width=0.95\columnwidth]{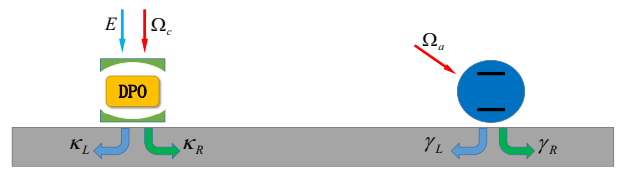}
\caption{(Color on line) Schematic diagram of the system. A two-level atom and a single-mode cavity containing a nonlinear medium are dissipatively coupled to a photonic waveguide and are driven by laser fields of Rabi frequencies $\Omega_{c}$ and $\Omega_{a}$. The nonlinear medium is pumped with the coupling strength $E$. }
\label{fig1}
\end{figure}

The system we consider consists of a two-level atom and a single-mode cavity coupled to a one-dimensional waveguide composed of a continuum of left and right propagating modes.
The atom has excited state $\ket e$ and ground state $\ket g$, and is damped at rates $\gamma_{R}$ and $\gamma_{L}$ into the guided modes propagating to the right and left directions, as illustrated in Fig.~\ref{fig1}.
The cavity mode is damped into the right and left guided modes at rates $\kappa_{R}$ and $\kappa_{L}$, respectively. The atom is driven by an external monochromatic field of frequency $\omega_{L}$, whereas the field driving the cavity is composed of two frequency components, i.e, $\omega_{p}$ pumping the nonlinear medium and $\omega_{L}$ driving the cavity mode. It is assumed that the pump and the field driving the cavity are tuned so that $\omega_{p}=2\omega_{L}$.  A single laser beam is used to provide the driving field for the atom and the cavity mode, and the pump for the nonlinear medium, after frequency doubling~\cite{wk86}. The pumped nonlinear medium serves as a DPO generating correlated pairs of photons of frequency $\omega_{L}$. The cavity is considered as a one-sided cavity, with one lossless input mirror and one output mirror of finite transmissivity located close to the waveguide.

We suppose that the decays of the cavity mode and the atom to the guided modes may exhibit a chirality that the decay to the right and to the left propagating modes can be asymmetric,
i.e., $\kappa_{R}\neq \kappa_{L}$ and $\gamma_{R}\neq\gamma_{L}$. The asymmetry of the decay is characterized by a parameter $\chi \in [0,1]$, defined as
\begin{align}
\chi = \frac{\kappa_{L}}{\kappa_{R}} = \frac{\gamma_{L}}{\gamma_{R}} ,\label{c1}
\end{align}
where $\chi =1$ corresponds to the symmetric (nondirectional) decay, $0<\chi <1$ corresponds to an asymmetric (partly directional) decay, and $\chi=0$ corresponds to the complete directional decay to only the right propagating modes.

To see the effects of the chirality of the guided modes on the dynamics of the system of atom plus cavity mode, we consider the master equation of the reduced density operator $\rho$ describing the cavity mode and the atom. The details of the derivation of the master equation are given in Appendix~\ref{App1}. The master equation is based on the Born-Markov approximations together with the one-dimensional approximation of mode structure of the waveguide field. If we choose $k_{0}x_{ac} =2\pi m,\, (m = 0,1,2,\ldots)$~\cite{rp14,pr15,cj12,gp13} where the distance $x_{ac}$ is commensurate
with the wavelength of the reservoir excitations, the master equation~(\ref{A17}) takes the form
\begin{align}
\dot{\rho} =& -i[\hat{H}^{\prime}_{sys},\rho]+\frac{1}{2}\kappa(\chi + 1)(2\hat{a}\rho \hat{a}^{\dag}-\rho \hat{a}^{\dag}\hat{a} -\hat{a}^{\dag}\hat{a}\rho) \nonumber\\
&+\frac{1}{2}\gamma(\chi + 1)(2\hat{\sigma}_{-}\rho \hat{\sigma}_{+}-\rho \hat{\sigma}_{+}\hat{\sigma}_{-}-\hat{\sigma}_{+}\hat{\sigma}_{-}\rho) \nonumber \\
&-\sqrt{\kappa\gamma}\left([\rho \hat{a}^{\dag},\hat{\sigma}_{-}] + [\hat{\sigma}_{+},\hat{a}\rho]\right) \nonumber\\
&- \chi\sqrt{\kappa\gamma}\left([\rho \hat{\sigma}_{+},\hat{a}] + [\hat{a}^{\dag},\hat{\sigma}_{-}\rho]\right) ,\label{c2}
\end{align}
where $\hat{\sigma}_{+}$ and $\hat{\sigma}_{-}$ are dipole raising and lowering operators for the atom, $\hat{a}^{\dagger}$ and $\hat{a}$ are creation and annihilation operators for the cavity mode, and $\kappa \equiv \kappa_{R}$ and $\gamma \equiv \gamma_{R}$. In Eq.~(\ref{c2}), $\hat{H}^{\prime}_{sys}$ is the Hamiltonian of the system,
\begin{align}
\hat{H}^{\prime}_{sys} &=  \Delta_{c}\hat{a}^{\dagger}\hat{a} +\Delta_{a}\hat{\sigma}_{+}\hat{\sigma}_{-} + \frac{1}{2}i\left(E^{\ast}\hat{a}^{2} - E \hat{a}^{\dagger 2}\right) \nonumber\\
& + i\left(\Omega_{c} \hat{a} +\Omega_{a} \hat{\sigma}_{-} - {\rm H.c.}\right) ,
\end{align}
where $\Delta_{c}=\omega_{c}-\omega_{L}, \Delta_{a}=\omega_{a}-\omega_{L}$ are the detuning of the cavity mode frequency $\omega_{c}$ and of the atomic resonance frequency $\omega_{a}$ from the driving field frequency $\omega_{L}$, $E$ is the coupling constant associated with the field pumping the nonlinear medium, and the quantities $\Omega_{c}$ and $\Omega_{a}$ are the Rabi frequencies associated with the external fields driving, respectively, the cavity mode and the atom. The Rabi frequencies have been chosen to be real and positive, whereas the coupling constant $E$ has been chosen to be a complex number, $E = |E|\exp(i\phi_{D})$, to allow for variations of the phase $\phi_{D}$ of the DPO field relative to the phase of the driving fields, $\phi_{c}=\phi_{a}=0$.

The dissipative part of the master equation (\ref{c2}) consists of terms describing damping of the cavity mode and the atom at rates $\kappa(\chi+1)/2$ and $\gamma(\chi+1)/2$, respectively, and also the cross coupling between the atom and the cavity mode mediated by the waveguide modes. The cross coupling reflects the fact that, as the atom decays to the guided modes, it drives the cavity mode, and vice versa.
Note the presence of the cross coupling terms only in the dissipative (incoherent) part of the master equation. There is no a cross coupling term present in the Hamiltonian (coherent) part of the master equation. In Appendix~\ref{App1}, we show that in general a term of this type is present [the term $\Omega_{ac}$ in Eq.~(\ref{A17})], but due to the specific choice of the distance $x_{ac}$, this term vanishes~\cite{rp14,pr15,cj12,gp13}.

\subsection{Convenient collective basis}

The presence of the dissipative cross coupling between the atom and the cavity mode suggests the introduction of superposition (polariton) operators such that
\begin{align}
\hat{J}^{\dagger} = u\hat{a}^{\dagger} +w\hat{\sigma}_{+} \quad {\rm and}\quad  \hat{B}^{\dagger} = w\hat{a}^{\dagger} - u\hat{\sigma}_{+} ,\label{c4}
\end{align}
and suggests description of the system in terms of superposition states, involving product states $\ket{g,n}\equiv \ket g\otimes\ket n$ and $\ket{e,n-1}\equiv \ket e \otimes\ket{n-1}$, where $\ket n$ is the photon number state of the field, and $\ket g$ and $\ket e$ are the atomic ground and excited states, respectively.

If we assume that the driving fields are weak, i.e., that $\Omega_{c},\Omega_{a}$, and $E$ are much smaller than the damping rates $\kappa$ and $\gamma$, we may consider the dynamics of the system in a Hilbert space  truncated at the $n=2$ manifold. In other words, we may assume that no more than two excitations are present in the system, i.e., we take into account the zero- $(n=0)$, one- $(n=1)$, and two-photon $(n=2)$ manifolds only. In this case, the Hilbert space is spanned by five state vectors defined as follows
\begin{align}
\ket 1 &\equiv \ket{g,0} ,\quad n=0 ,\nonumber\\
\ket 2 &\equiv \ket{g,1} ,\quad \ket 4 \equiv \ket{e,0} ,\quad n=1,\nonumber\\
\ket 3 &\equiv \ket{g,2} ,\quad \ket 5 \equiv \ket{e,1} ,\quad n=2 .\label{c8a}
\end{align}

Before going into a detailed analysis of the dynamics between the superposition states, let us first determine the role of the $\hat{J}^{\dagger}$ and $\hat{B}^{\dagger}$ operators. We make use of the representations of the superposition operators in the basis of the product states (\ref{c8a}), which are given in Appendix~\ref{App2}. From Eq.~(\ref{A2}) we readily find that the operators $\hat{J}^{\dagger}$ and $\hat{B}^{\dagger}$  project the ground state $\ket 1$ to single-excitation states
\begin{align}
\hat{J}^{\dagger}\ket{1} = \ket{\psi} \quad {\rm and}\quad \hat{B}^{\dagger}\ket{1} = \ket{\phi} ,
\end{align}
where
\begin{align}
\ket\psi = u\ket{2} + w\ket{4} ,\quad \ket\phi = w\ket{2} -u\ket{4} ,
\end{align}
are symmetric and antisymmetric superpositions of the single-excitation states, and $u^{2}+w^{2}=1$. This shows that the $\hat{J}^{\dagger}$ and $\hat{B}^{\dagger}$ operators are the raising operators of the  $\ket{1}\leftrightarrow\ket{\psi}$ and  $\ket{1}\leftrightarrow\ket{\phi}$ transitions, respectively.

Similarly, the action of the superposition operators on the single-excitation states projects the states to double-excitation states
\begin{align}
\hat{J}^{\dagger}\ket{\psi} &= \sqrt{2}u\left(u\ket{3} +\sqrt{2}w\ket{5}\right) ,\label{c7}\\
\hat{B}^{\dagger}\ket{\phi} &= \sqrt{2}w\left(w\ket{3} -\sqrt{2}u\ket{5}\right) .\label{c8}
\end{align}
It is easily verified that the resulting superpositions of the double-excitation states appearing on the right-hand side of Eqs.~(\ref{c7}) and (\ref{c8}) are not orthogonal. However, the superpositions can be expressed in terms of orthogonal states
\begin{align}
\hat{J}^{\dagger}\ket{\psi} &= \sqrt{2}u\!\left[\!\left(\!u\alpha\!+\!\sqrt{2}w\beta\right)\!\ket\xi\!+\!\left(\!u\beta\!-\!\sqrt{2}w\alpha\right)\!\ket\zeta\!\right] ,\label{c9}\\
\hat{B}^{\dagger}\ket{\phi} &= \sqrt{2}w\!\left[\!\left(\!w\alpha\!-\!\sqrt{2}u\beta\right)\!\ket\xi\!+\!\left(\!w\beta\!+\!\sqrt{2}u\alpha\right)\!\ket\zeta\!\right] ,\label{c10}
\end{align}
where
\begin{align}
\ket\xi &= \alpha \ket{3} +\beta \ket{5} ,\quad \ket\zeta = \beta\ket{3} -\alpha\ket{5} \label{c11}
\end{align}
are symmetric and antisymmetric superpositions of the double-excitation states, and $\alpha^{2}+\beta^{2}=1$.
It follows that the raising operator $\hat{J}^{\dagger}$ projects the system from the state $\ket\psi$ and the raising operator $\hat{B}^{\dagger}$ projects the state $\ket\phi$ to two double-excitation states $\ket\xi$ and~$\ket\zeta$. One can notice from Eqs.~(\ref{c9}) and (\ref{c10}) that depending on the choice of the parameters either $\hat{J}^{\dagger}$ can project $\ket\psi$ to the state $\ket\xi$ only or $\hat{B}^{\dagger}$ can project $\ket\phi$ to the state $\ket\zeta$ only. That is, we require $u\beta =\sqrt{2}w\alpha$ for $\hat{J}^{\dagger}$ to project $\ket\psi$ to the state $\ket\xi$ only. Similarly, a choice $w\alpha =\sqrt{2}u\beta$ results in $\hat{B} ^{\dagger}$ to project $\ket\phi$ to the state $\ket\zeta$ only.

We would like to point out that the superposition states $\ket\psi,\ket\phi$ and $\ket\xi,\ket\zeta$ are defined in terms of arbitrary parameters $u,w$ and $\alpha,\beta$, respectively. However, this arbitrariness will not affect our results, and the parameters can be chosen by the dictates of convenience for specific calculations. If we choose
\begin{align}
u = \sqrt{\frac{\kappa}{\kappa +\gamma}} ,\quad w = \sqrt{\frac{\gamma}{\kappa +\gamma}} ,\label{c10a}
\end{align}
we can rewrite the master equation (\ref{c2}) in terms of the superposition operators as
\begin{align}
\dot{\rho} = -i\left[\hat{H}_{e},\rho\right] +\frac{1}{2}\Gamma_{\chi}\left(2\hat{J}\rho \hat{J}^{\dagger} -\rho \hat{J}^{\dagger}\hat{J} -\hat{J}^{\dagger}\hat{J}\rho\right) ,\label{c11}
\end{align}
where we have used
\begin{align}
\hat{H}_{e} &= \left[\Delta_{s}-\left(u^{2}-w^{2}\right)\Delta\right]\hat{B}^{\dagger}\hat{B} \nonumber\\
&+ \left[\Delta_{s} + \left(u^{2}-w^{2}\right)\!\Delta\right]\hat{J}^{\dagger}\hat{J} +2uw\Delta\!\left(\hat{B}^{\dagger}\hat{J} + \hat{J}^{\dagger}\hat{B}\right) \nonumber\\
&+ \frac{1}{2}i\left[E^{\ast}\!\left(w \hat{B} +u \hat{J}\right)^{2} - {\rm H.c.}\right] \nonumber\\
&+ i\!\left(\Omega_{\psi}\hat{J} + \Omega_{\phi}\hat{B} - {\rm H.c.}\right) + iG_{\chi}\!\left(\hat{B}^{\dagger}\hat{J} - \hat{J}^{\dagger}\hat{B}\right) ,\label{c4a}
\end{align}
with
\begin{align}
\Delta_{s} &\equiv \frac{1}{2}(\Delta_{c}+\Delta_{a}) ,\quad \Delta\equiv \frac{1}{2} (\Delta_{c}-\Delta_{a}) ,\nonumber\\
G_{\chi} &\equiv \frac{1}{2}(1-\chi)\sqrt{\kappa\gamma} ,\quad \Gamma_{\chi}\equiv (1+\chi)(\kappa+\gamma) ,\nonumber\\
\Omega_{\psi} &\equiv  \left(u\Omega_{c} + w\Omega_{a}\right) ,\quad \Omega_{\phi}\equiv \left(w\Omega_{c} -u\Omega_{a}\right) .
\end{align}
It is seen from the structure of the master equation that the chirality has been completely incorporated into the effective Hamiltonian of the system. Note that the chiral part of the Hamiltonian is Hermitian. The chirality coupling parameter $G_{\chi}$ introduces a coupling between the symmetric and antisymmetric states of the system. The parameter $\Delta$ introduces frequency shifts and also plays a role of a coupling between the the symmetric and antisymmetric states. This coupling vanishes when the driving field frequency is on resonance with the atomic and the cavity mode frequencies $(\Delta_{a}=\Delta_{c}=0)$, or is equally detuned from the resonances, $\Delta_{a}=\Delta_{c}$.

The energies of the symmetric and antisymmetric states are shifted from $\Delta_{s}$ by an amount $(u^{2}-w^{2})\Delta$. This shows that their degeneracy is lifted only in the non-symmetric case of $u\neq w$ and $\Delta_{c}\neq \Delta_{a}$. Since $u\neq w$ implies $\kappa\neq \gamma$, and $\Delta_{c}\neq \Delta_{a}$ implies $\omega_{c}\neq \omega_{a}$, the shift is present only for systems of unequal damping rates and unequal angular frequencies. The symmetric superposition is coupled to the driving field with an enhanced Rabi frequency $\Omega_{\psi}$, whereas the antisymmetric superposition is coupled to the driving field with a reduced Rabi frequency~$\Omega_{\phi}$.

The dissipative part of the master equation is determined solely by the $\hat{J}^{\dagger}$ and $\hat{J}$ operators which are the raising and lowering operators of the $\ket{1}\leftrightarrow\ket{\psi}$ and $\ket{\psi}\leftrightarrow\ket{\xi},\ket\zeta$ transitions. This means that the antisymmetric state $\ket\phi$ does not decay at all, although it is directly coupled to the ground state by the driving field with the Rabi frequency~$\Omega_{\phi}$.

\subsection{Spontaneous transition channels}

Consider first the contribution of the dissipative part of the master equation (\ref{c11}) to the equations of motion for the diagonal elements of the density operator, i.e., the rate equations which correspond to the evolution of the populations of the states. This will give us information about transition rates between the collective states of the system.

If we consider only the dissipative part of the master equation, the diagonal density matrix elements, which correspond to the populations of the superposition states, evolve as
\begin{align}
\dot{\rho}_{\phi\phi} =&\, \Gamma_{\xi\phi}\rho_{\xi\xi} +\Gamma_{\zeta\phi}\rho_{\zeta\zeta} + \sqrt{\Gamma_{\xi\phi}\Gamma_{\zeta\phi}}\left(\rho_{\xi\zeta}+\rho_{\zeta\xi}\right)  ,\label{c16}\\
\dot{\rho}_{\psi\psi} =& -\Gamma_{\chi}\rho_{\psi\psi} + \Gamma_{\xi\phi}\rho_{\xi\xi} +\Gamma_{\zeta\psi}\rho_{\zeta\zeta} \nonumber\\
&+ \sqrt{\Gamma_{\xi\psi}\Gamma_{\zeta\psi}}\left(\rho_{\xi\zeta}+\rho_{\zeta\xi}\right) , \label{c17}
\end{align}
and
\begin{align}
\dot{\rho}_{\xi\xi} =& -\left(\Gamma_{\xi\phi} + \Gamma_{\xi\psi}\right)\rho_{\xi\xi} \nonumber\\
&-\frac{1}{2}\!\left(\sqrt{\Gamma_{\xi\phi}\Gamma_{\zeta\phi}} +\sqrt{\Gamma_{\xi\psi}\Gamma_{\zeta\psi}}\right)\!\left(\rho_{\xi\zeta}+\rho_{\zeta\xi}\right) ,\label{c19}\\
\dot{\rho}_{\zeta\zeta} =& -\left(\Gamma_{\zeta\phi} + \Gamma_{\zeta\psi}\right)\rho_{\zeta\zeta} \nonumber\\
&-\frac{1}{2}\!\left(\sqrt{\Gamma_{\xi\phi}\Gamma_{\zeta\phi}} +\sqrt{\Gamma_{\xi\psi}\Gamma_{\zeta\psi}}\right)\!\left(\rho_{\xi\zeta}+\rho_{\zeta\xi}\right) ,\label{c20}
\end{align}
where
\begin{align}
\Gamma_{\xi\phi} &= \Gamma_{\chi}\!\left(\sqrt{2}w\eta -\beta\right)^{2} ,\quad
\Gamma_{\xi\psi} = 2u^{2}\Gamma_{\chi}\eta^{2} ,\nonumber\\
\Gamma_{\zeta\phi} &= \Gamma_{\chi}\left(\sqrt{2}w\sigma +\alpha\right)^{2} ,\quad
\Gamma_{\zeta\psi} = 2u^{2}\Gamma_{\chi}\sigma^{2} ,
\end{align}
are the damping rates of the transitions between the collective double- and single-excited states with $\eta = \alpha u +\sqrt{2}w\beta$ and $\sigma =\beta u -\sqrt{2}w\alpha$.

First of all, we note that the population of the state $\ket\phi$ does not decay, but can be populated by spontaneous emission from the double-excitation states $\ket\xi$ and $\ket\zeta$.
The population of the state $\ket\psi$ decays to the ground state with a rate $\kappa +\gamma$ and is populated by spontaneous decay from the $\ket\xi$ and $\ket\zeta$ states.
Both of the double-excitation states decay to the $\ket\psi$ and $\ket\phi$ states, but then only a part of the population, i.e., that one in the state $\ket\psi$, can decay to the ground state. Thus, a part of the population can be trapped in the state $\ket\phi$. Consequently, the steady state of the system may not be the ground state, but rather an incoherent mixture of the single excitation $\ket\phi$ and the ground~$\ket 1$ states, ${\rm Tr}\rho^{2}=\rho_{11}^{2}+\rho_{\phi\phi}^{2} < 1$.

\subsection{Effect of the chiral term of the Hamiltonian}

\begin{figure}[h]
\centering\includegraphics[width=0.75\columnwidth]{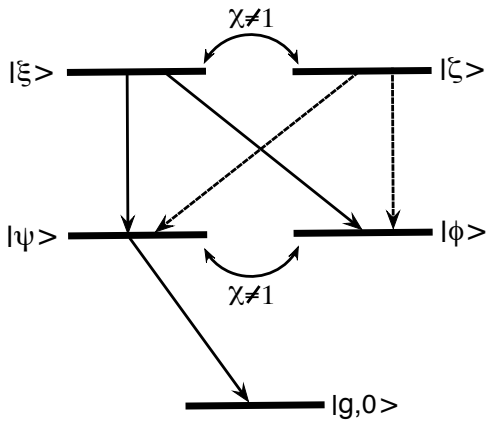}
\caption{Energy level structure and the allowed spontaneous transitions channels of the undriven system $(E= \Omega_{c} =\Omega_{a} =0)$. The presence of the chirality $(\chi\neq 1)$ results in a coherent coupling between the superposition states. }
\label{fig2}
\end{figure}

We now turn to determine the contribution of the chiral term in Eq.~(\ref{c4a}) to the evolution of the density matrix elements.
The effect of having the chiral term in the Hamiltonian of the system is seen from Eq.~(\ref{c4a}) to be a coupling between the $\ket\psi$ and $\ket\phi$ states. The coupling may lead to a transfer of population between these states. It can be seen more clearly if one considers the contribution of the chiral term to the equations of motion for coherences between the excited states. It is easily verified that the chiral term contributes to the equation of motion for the coherence $\rho_{\psi\phi}$, which is in the form
\begin{align}
\dot{\rho}_{\psi\phi} = G_{\chi}\left(\rho_{\psi\psi} - \rho_{\phi\phi}\right) .
\end{align}
Clearly, the chiral term gives rise to coherence between the superposition states. Note that the equation of motion for the coherence is independent of $u$ and $w$. Moreover, the form of the equation is the same as that for a two-level system driven by a coherent field. It follows then that the chirality appears as a coherent field driving the $\ket\psi\leftrightarrow \ket\phi$ transition with the coupling constant $G_{\chi}$ playing a role of the Rabi frequency of the field, as illustrated in Fig.~\ref{fig2}.

Similarly, we may calculate the evolution of the coherence between the two-photon superpositions states, and find that the chiral term makes the contribution
\begin{align}
\dot{\rho}_{\xi\zeta} = \sqrt{2}G_{\chi}\left(\rho_{\xi\xi} - \rho_{\zeta\zeta}\right) .
\end{align}
We first observe that the contribution of the chiral term to the coherence is independent of $\alpha$ and $\beta$.
As in the case of the single-excitation states, the contribution of the chiral term is analogous to that of a coherent field of the Rabi frequency $\sqrt{2}G_{\chi}$ driving the $\ket\xi\leftrightarrow \ket\zeta$ transition.

We may conclude that in the coupled systems problem that we are considering the most obvious effect of having chirality $(\chi\neq 1)$ can be seen as a coherent coupling between the energy states of the same number of excitations. The chirality provides another route to populate the excited states that it can transfer the population between the excited states leading, for example, to an increased population of the $\ket\phi$ state at the expense of a decreased population of the $\ket\psi$ state.

The contribution of the remaining terms of Eq.~(\ref{c4a}) to the equations of motion for the density matrix elements can be easily evaluated, and the complete set of the equations of motion is given in Appendix~\ref{App3}.

We finish this section with a brief discussion of the parameters characterizing the driven DPO cavity, i.e., the atom and the photonics waveguide, and the ranges of these parameters experimentally accessible. Driving lasers used in experiments are usually tunable, providing for arbitrary detunings $\Delta_{a}$. Cavity frequencies can be easily controlled using, for example, piezoelectric transducers so that the cavity detuning $\Delta_{c}$ can be arbitrarily varied. The crucial factor is to achieve a strong coupling efficiency of the cavity and the atom to the photonic waveguide. Recent experiments have demonstrated that single atoms as well as quantum dots can be strongly coupled to the waveguide with near perfect efficiency~\cite{ll07,jo13,as14}.

\section{Decoherence-free subspace and dark states}\label{sec3}

Having explained the physical processes involved in the dynamics of the system we now turn to determine if the system could evolve to a stationary state $(\dot{\rho}=0)$, which is a dark state defined by the following conditions~\cite{bb00,lw03}:
\begin{align}
&(a)\quad {\cal L}\rho\equiv  \frac{1}{2}\Gamma_{\chi}\left(2\hat{J}\rho \hat{J}^{\dagger} -\rho \hat{J}^{\dagger}\hat{J} -\hat{J}^{\dagger}\hat{J}\rho\right) = 0 ,\nonumber\\
&(b)\quad \hat{J}\left(\hat{H}_{e}\rho\right) = 0 ,\nonumber\\
&(c)\quad \hat{H}_{e}\rho = 0 .\label{c26}
\end{align}
Since ${\cal L}\rho$ contains all the terms responsible for decoherence, the condition $(a)$ ensures that the stationary state belongs to a subspace which is decoherence-free.
The condition $(b)$ ensures that the unitary evolution under $\hat{H}_{e}$ does not take the state out of the decoherence-free subspace. Finally, the condition $(c)$ ensures
the state which belongs to the decoherence-free subspace does not evolve. In other words, the condition $(c)$ implies that the stationary state is a {\it dark state} which does not evolve at all, i.e., the state is completely decoupled from the remaining states.

Note that the dissipation part of the master equation involves only the $\hat{J}$ and $\hat{J}^{\dagger}$ operators. Hence, if the steady state exists as a pure state $(\rho =\ket\Psi\bra\Psi)$, we then can replace the condition $(a)$, ${\cal L}\rho =0$, by a simplified condition $\hat{J}\ket{\Psi}=0$.

Before going into detailed analysis of the stationary state of the system, it is important to understand the role the chirality will play in the creation of decoherence-free and dark states.
We have already noticed that the chirality is completely contained in the Hamiltonian $\hat{H}_{e}$. Then, it is clear from Eq.~(\ref{c26}) that the system can decay to a decoherence-free state independent of whether or not the chirality is present. Evidently, the role of the chirality is to ensure that the decoherence-free state remains inside the decoherence-free subspace and that under additional conditions, it could become a dark state.

We now proceed to examine the stationary state of our system. We are particularly interested in the role of the chirality in the decay of the system to a dark state. There are two cases that we shall consider.
In the first, we shall limit our analysis to the Hilbert space truncated at the single-excitation states corresponding to $n\leq 1$. In the second, we extend the analysis to include the double excitation states. Although the truncation of the Hilbert space is an approximation, it offers some advantages over the exact solutions. Because of the simplicity, we shall obtain detailed and almost exact analytical solutions. The validity of the truncation is then tested by numerically solving the master equation in the complete Hilbert space of the system.

\subsection{Single excitation in the system}

Let us first examine if one could obtain a pure decoherence-free state, different from the ground state, by employing the weak field assumption and deciding to only consider the zero- and one-photon states. In the case in which there could be only one excitation present in the system the Hilbert space of the system is spanned by three state vectors, $\{\ket 1, \ket\phi,\ket\psi\}$.

To examine the occurrence of a decoherence-free subspace, we must look at conditions (\ref{c26}). Consider a linear superposition
\begin{align}
\ket{\Psi_{1}} = c_{1}\ket 1 +c_{\phi}\ket\phi +c_{\psi}\ket\psi ,
\end{align}
where the subscript $"1"$ refers to single-excitation states only.
The condition ${\cal L}\rho =0$ implies
\begin{align}
\hat{J}\ket{\Psi_{1}} = 0 .
\end{align}
Since $\hat{J}=\ket{1}\bra{\psi}$, we readily find that $c_{\psi}=0$. This implies that the subspace which is free of spontaneous emission is spanned by two state vectors $\ket 1$ and $\ket\phi$.

We now check the remaining condition for a decoherence free space where the vector $\hat{H}_{e}\ket{\Psi_{1}}$ remains in the subspace $\{\ket 1,\ket\phi\}$. We can write this condition as
\begin{align}
\hat{J}\left(\hat{H}_{e}\ket{\Psi_{1}}\right) = 0 .\label{c41}
\end{align}

Applying the Hamiltonian on the state $\ket{\Psi_{1}}$, we obtain
\begin{align}
\hat{H}_{e}\ket{\Psi_{1}} &= i\Omega_{\phi}c_{\phi}\ket 1 \nonumber\\
&+ \left\{[\Delta_{s} - (u^{2} -w^{2})\Delta]c_{\phi} -i\Omega_{\phi}c_{1}\right\}\ket\phi \nonumber\\
&+ \left[\left(2uw\Delta - iG_{\chi}\right)c_{\phi} -i\Omega_{\psi}c_{1}\right]\ket\psi ,\label{c42}
\end{align}
from which we readily find that the condition (\ref{c41}) is satisfied~if
\begin{align}
c_{\phi} = \frac{i\Omega_{\psi}\left(2uw\Delta + iG_{\chi}\right)}{\left(2uw\Delta\right)^{2}+G_{\chi}^{2}}\, c_{1} .\label{c43}
\end{align}
Hence, the decoherence-free state is of the form
\begin{align}
\ket{\Psi_{1}} = \frac{\sqrt{(2uw\Delta)^{2}+G_{\chi}^{2}}\ket 1  +i\Omega_{\psi}e^{i\theta}\ket\phi}{\sqrt{(2uw\Delta)^{2}+G_{\chi}^{2}+\Omega_{\psi}^{2}}} ,\label{c44}
\end{align}
where $\theta =\arctan(G_{\chi}/2uw\Delta)$.

This example demonstrates the possibility of generating a decoherence free subspace of the system by the chiral interaction $(G_{\chi}\neq 0)$ as well as by unequal detunings $\Delta_{c}\neq \Delta_{a}$ of the driving fields. In both cases, the decoherence free state produced involves only the the single-excitation state $\ket\phi$ and the ground state $\ket 1$.

We stress that the state (\ref{c44}) undergoes a dynamical evolution since, according to Eq.~(\ref{c42}), $\hat{H}_{e}\ket{\Psi_{1}}\neq 0$. For $\hat{H}_{e}\ket{\Psi_{1}}= 0$, it is required that all of the coefficients appearing in Eq.~(\ref{c42}) are zero where, in addition to the condition (\ref{c43}), it is required that
\begin{align}
\Omega_{\phi}=0 \quad {\rm and}\quad \left[\Delta_{s}-\left(u^{2}-w^{2}\right)\Delta\right] = 0 .\label{c45}
\end{align}
Under the conditions (\ref{c43}) and (\ref{c45}), the state $\ket{\Psi_{1}}$ does not evolve when it is completely decoupled from the dynamics of the system. In other words, the state $\ket{\Psi_{1}}$ becomes a dark state. Note that the conditions~(\ref{c45}) are fulfilled when $\Omega_{c}/\Omega_{a}=u/w$ and $\Delta_{c}/\Delta_{a} = -(u/w)^{2}$.

It is easily verified from the Hamiltonian (\ref{c4a}) that the condition $\Omega_{\phi}=0$ is related physically to assume that the transition $\ket 1 \rightarrow \ket\phi$ is not directly driven by the laser field, and the other condition is related physically to the assumption that the driving fields are on resonance with the $\ket 1 \rightarrow \ket\phi$ transition.

\begin{figure}[h]
\centering\includegraphics[width=0.95\columnwidth]{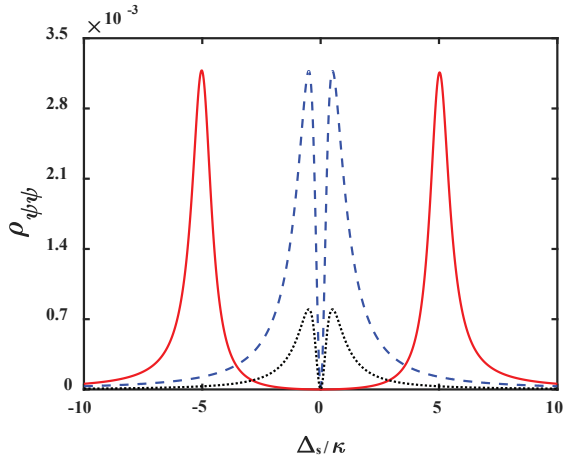}
\caption{(Color online) The stationary population of the state $\ket\psi$ plotted as a function of $\Delta_{s}/\kappa$ for $\gamma=\kappa\, (u=w)$, $G_{\chi}=5\kappa$, and several different values of $\Delta$ and $\Omega\equiv \Omega_{c}=\Omega_{a}$: $\Delta=5\kappa, \Omega =0.04\kappa$ (solid red line), $\Delta=0, \Omega =0.04\kappa$ (dashed blue line), and $\Delta=0, \Omega =0.02\kappa$ (dotted black line).}
\label{fig3}
\end{figure}

One can notice that among the conditions imposed on the parameters to create the dark state (\ref{c44}), there is no condition imposed on the Rabi frequency $\Omega_{\psi}$. This means that the creation of the dark state leaves the symmetric state $\ket\psi$ unpopulated despite the fact that it is continuously driven by the laser field. This feature is easily seen in Fig.~\ref{fig3}, which shows the population $\rho_{\psi\psi}$ as a function of $\Delta_{s}$ for several different values of~$\Omega_{\psi}$. It is apparent that $\rho_{\psi\psi}=0$ at $\Delta_{s}=0$ regardless of the value of $\Omega_{\psi}$.

Finally, we would like to point out an interesting effect of the chirality on the dark states of the system. One can see from Eq.~(\ref{c44}) that in the non-chiral situation, $G_{\chi}=0$, and $\Delta=0$, the dark state $\ket{\Psi_{1}}$ reduces to $\ket\phi$. However, if the chirality is present $(G_{\chi}\neq 0)$, the state $\ket\phi$ is no longer the dark state but rather a linear superposition of the ground $\ket g$ and $\ket\phi$ state.
In this situation the system behaves as an effective two-level system.

\subsection{Double excitation in the system}

In the previous section, we demonstrated that the chiral interaction $(G_{\chi}\neq 0)$ between a driven two-level atom $(\Omega_{a}\neq 0)$ and a driven empty cavity $(\Omega_{c}\neq 0, E=0)$ can produce a dark state in the system which is a linear combination of the asymmetric single excitation state $\ket\phi$ and the ground state $\ket 1$. In this section, we demonstrate that the inclusion of the double-excitation states may result in the system decaying to the same dark state, as in the $n=1$ case, if one replaces the driven empty cavity with a DPO, $(E\neq 0)$.

When up to two excitations are present, the Hilbert space of the system is spanned by five state vectors $\{\ket 1,\ket\psi, \ket\phi, \ket\xi, \ket\zeta\}$.
In this case, the calculation of the conditions for the decoherence-free subspace and the production of a dark state is considerably more complicated than the calculation in the $n=1$ case, since we must include
more basis states than the three used to derive Eq.~(\ref{c44}). The $\hat{J}$ operator is modified by the presence of the double-excitation states and has the form
\begin{align}
\hat{J} &= \ket{1}\bra{\psi} +\sqrt{2}\left(u\ket\psi +w\ket\phi\right)\left(\eta\bra\xi + \sigma\bra\zeta\right) \nonumber\\
&+\ket\phi\left(\alpha\bra\zeta -\beta\bra\xi\right) .\label{c47}
\end{align}

Let us introduce a superposition state
\begin{align}
\ket{\Psi_{2}} = c_{1}\ket 1 +c_{\phi}\ket\phi +c_{\psi}\ket\psi +c_{\xi}\ket\xi +c_{\zeta}\ket\zeta ,\label{c47a}
\end{align}
where the subscript $"2"$ indicates the inclusion of the double-excitation states. If we apply $\hat{J}$ on the state $\ket{\Psi_{2}}$, we readily find that the condition $\hat{J}\ket{\Psi_{2}}=0$ is satisfied
when $c_{\psi}=c_{\xi}=c_{\zeta} = 0$. Like in the single-excitation case, the state which is free of spontaneous emission is again spanned by the two state vectors $\ket 1$ and~$\ket\phi$.

In order to check the remaining condition for the decoherence-free space, we evaluate the vector $\hat{H}_{e}\ket{\Psi_{2}}$ and find
\begin{align}
\hat{H}_{e}\ket{\Psi_{2}} &= i\Omega_{\phi}c_{\phi}\ket 1 \nonumber\\
&+ \left\{[\Delta_{s} - (u^{2} -w^{2})\Delta]c_{\phi} -i\Omega_{\phi}c_{1}\right\}\ket\phi \nonumber\\
&+ \left[\left(2uw\Delta - iG_{\chi}\right)c_{\phi} -i\Omega_{\psi}c_{1}\right]\ket\psi \nonumber\\
&+i\left\{-\frac{\alpha E}{\sqrt{2}} c_{1}-\left(w\beta\Omega_{a} -\sigma\Omega_{c}\right)c_{\phi}\right\}\ket\xi \nonumber\\
&+i\left\{-\frac{\beta E}{\sqrt{2}} c_{1}+\left(w\alpha\Omega_{a} -\eta\Omega_{c}\right)c_{\phi}\right\}\ket\zeta .\label{c48}
\end{align}
Under the action of $\hat{J}$ the vector $\hat{H}_{e}\ket{\Psi_{2}}$ evolves into the state
\begin{align}
\hat{J}\left(\hat{H}_{e}\ket{\Psi_{2}}\right) &= \left[\left(2uw\Delta - iG_{\chi}\right)c_{\phi} -i\Omega_{\psi}c_{1}\right]\ket{1} \nonumber\\
&-i\left\{uwEc_{1}+\left[u\Omega_{c}-w\!\left(u^{2}\!-\!w^{2}\right)\!\Omega_{a}\right]\!c_{\phi}\right\}\!\ket\phi \nonumber\\
&-iu\left(uEc_{1} +2w^{2}\Omega_{a}c_{\phi}\right)\ket\psi .\label{c49}
\end{align}
From Eq.~(\ref{c49}) it is apparent by inspection that the result $\hat{J}(\hat{H}_{e}\ket{\Psi_{2}}) = 0$ is impossible unless we set the Rabi frequencies to satisfy the condition $u\Omega_{c}=w\Omega_{a}$. Under this condition, Eq.~(\ref{c49}) assumes the simplified form
\begin{align}
\hat{J}\left(\hat{H}_{e}\ket{\Psi_{2}}\right) &= \left[\left(2uw\Delta - iG_{\chi}\right)c_{\phi} -2iu\Omega_{c}c_{1}\right]\ket{1} \nonumber\\
&-iuw\left(Ec_{1}+2w\Omega_{c}c_{\phi}\right)\ket\phi \nonumber\\
&-iu^{2}\left(Ec_{1}+2w\Omega_{c}c_{\phi}\right)\ket\psi .\label{c50}
\end{align}
From Eq.~(\ref{c50}) it then follows that the state $\hat{H}_{e}\ket{\Psi_{2}}$ remains in the subspace $\{\ket 1,\ket\phi\}$ if
\begin{align}
c_{\phi} &= \frac{2u\Omega_{c}e^{i(\theta+\frac{\pi}{2})}}{\sqrt{(2uw\Delta)^{2} + G^{2}_{\chi}}}\, c_{1} ,\label{c51a}\\
E &= \frac{4uw\Omega_{c}^{2}e^{i(\theta-\frac{\pi}{2})}}{\sqrt{(2uw\Delta)^{2} + G^{2}_{\chi}}} .\label{c51b}
\end{align}
Note the fact that the result for $\hat{J}(\hat{H}_{e}\ket{\Psi_{2}}) = 0$ holds irrespective of the choice of $\alpha$ and $\beta$, and is valid for $u=w$ as well as for $u\neq w$. The requirement that $E\neq 0$
demonstrates the importance of the DPO in the production of the decoherence-free subspace, i.e., it would not be possible to produce the decoherence-free subspace if the DPO were absent.

To examine if the state $\ket{\Psi_{2}}$ could be a dark state, we must look at the condition $\hat{H}_{e}\ket{\Psi_{2}} = 0$. It is easy to see from Eq.~(\ref{c48}) that $\ket{\Psi_{2}}$ becomes a dark state when, in addition to the conditions specified in Eqs.~(\ref{c51a}) and (\ref{c51b}),
\begin{align}
\Omega_{\phi} &= 0 ,\nonumber\\
\Delta_{s} - (u^{2} -w^{2})\Delta &= 0 ,
\end{align}
Note that the condition $\Omega_{\phi}=0$ required for $\ket{\Psi_{2}}$ to be a dark state actually imposes an additional constrain on the Rabi frequencies that $w\Omega_{c}=u\Omega_{a}$. This constraint together with the already established constraint $u\Omega_{c}=w\Omega_{a}$ imposes a limit on the coefficients $u$ and $w$ that $u=w$ should hold, i.e., damping rates of the cavity mode and the atom should be equal, $\kappa =\gamma$.
Hence, under this constraint, the conditions for the state $\ket{\Psi_{2}}$ to be a dark state become particularly simple and have the form
\begin{align}
\Omega_{a} &= \Omega_{c} ,\quad \kappa =\gamma ,\quad \Delta_{s}=0 ,\nonumber\\
c_{\phi} &= \frac{\sqrt{2}i\Omega_{c}e^{i\theta}}{\sqrt{\Delta^{2} + G^{2}_{\chi}}}c_{1} ,\quad
E = \frac{2\Omega_{c}^{2}e^{i(\theta-\frac{\pi}{2})}}{\sqrt{\Delta^{2} + G^{2}_{\chi}}} .\label{c53}
\end{align}
Evidently, the production of the dark state in the system depends strongly on the chirality. It is seen particularly well when $\Delta =0$. In this case, the chirality $(G_{\chi}\neq 0)$ is the only mechanism which produces the dark state. In the experimental situation, if there is some nonzero population present in the states $\ket\psi, \ket\xi$, or $\ket\zeta$, then fluorescence will be detected; if not, then the system will be in the dark state.

We emphasize that the amplitude $E$ depends on the phase $\theta-\pi/2$, which for a nonzero $\Delta$ is different from zero. Thus, a nonzero detuning $\Delta$ introduces a shift of the phase of the DPO.
To see it, we write the coupling strength $E =|E|\exp(i\phi_{D})$, where $\phi_{D}$ is the phase of the DPO field, and find from Eq.~(\ref{c51b}) that $|E|$ is real and positive at $\phi_{D} = \theta-\pi/2$.

To confirm that under the conditions (\ref{c53}) the system decays to a pure state, we find the steady-state values of the density matrix elements by
solving the equations of motion, Eqs.~(\ref{B1})$-$(\ref{B3}) in the limit of $t\rightarrow\infty$. The general solution of these equations is quite lengthy. Therefore, for clarity, we present analytical expressions for the steady-state values of only these density matrix elements which are nonzero under the conditions~(\ref{c53}). The expressions are
\begin{align}
\rho_{11} &= \frac{G^{2}_{\chi}+\Delta^{2}}{G^{2}_{\chi}+\Delta^{2}+2\Omega_{c}^{2}} ,\quad
\rho_{\phi\phi} = \frac{2\Omega^{2}_{c}}{G^{2}_{\chi}+\Delta^{2}+2\Omega_{c}^{2}} ,\nonumber\\
\rho_{1\phi}&=\frac{\sqrt{2}\Omega_{c}\sqrt{G^{2}_{\chi}+\Delta^{2}}}{G^{2}_{\chi}+\Delta^{2}+2\Omega^{2}_{c}}e^{-i(\theta+\frac{\pi}{2})} .
\end{align}
Clearly, only the populations of the ground $\ket 1$ and the single-photon $\ket\phi$ states, and coherence between them, are different from zero. Then it is easily to check that
\begin{align}
{\rm Tr}\rho^{2} = \rho_{11}^{2} +\rho_{\phi\phi}^{2} +2|\rho_{1\phi}|^{2} =1 .
\end{align}
Thus, under conditions (\ref{c53}) the stationary state of the system is the pure dark state.

\section{Photon blockade effect}\label{sec4}

The conditions (\ref{c53}) for the production of the dark state, which is a superposition of only the single excitation and the ground states, are a clear implication of the presence of the photon blockade effect in the system. The photon blockade prevents absorption of more than one photon by an optical system. This shows that there is a connection between the generation of the dark state $\ket{\Psi_{2}}$ and the photon blockade effect.

However, the photon blockade mechanism in our system is different from the conventional photon blockade. The conventional photon blockade requires either a Kerr nonlinearity in the system~\cite{tl94,is97,bi11,hs11,kk16}, or tuning all the double-excitation states off-resonant with the driving fields. The off-resonant tuning is usually achieved through a strong coherent coupling of an emitter to the field resulting in a dynamic Stark splitting of the energy states~\cite{tc92,bb05}. In our system the photon blockade is achieved with the double-excitation states on resonance with the driving fields and the emitters dissipatively (incoherently) coupled to the field. In this case, the photon blockade results from a destructive interference between different pathways from the ground state $\ket 1$ to the double-excited states $\ket\xi$ and $\ket\zeta$.

The double-excitation states are prevented from being populated due to the distractive quantum interference between
two alternative excitation pathways, namely, via stepwise transitions $\ket 1\rightarrow\ket\psi\rightarrow\ket\xi$, $\ket 1\rightarrow\ket\psi\rightarrow\ket\zeta$ with the driving fields $(\Omega_{c},\Omega_{a})$ and via two-photon transitions $\ket 1\rightarrow \ket\xi$, $\ket 1\rightarrow \ket\zeta$ with the two-photon field $E$. With the help of  the Schr\"{o}dinger equation $i\partial \ket{\Psi_{2}}/\partial t = \hat{H}_{e}\ket{\Psi_{2}}$, using Eqs.~(\ref{c48}) and (\ref{c51a}), we readily find
\begin{align}
\dot{c}_{\xi} &= -\frac{\alpha}{\sqrt{2}}\left[E + \frac{4uw\Omega_{c}^{2}e^{i(\theta+\frac{\pi}{2})}}{\sqrt{(2uw\Delta)^{2} + G^{2}_{\chi}}}\right]c_{1} ,\nonumber\\
\dot{c}_{\zeta} &= -\frac{\beta}{\sqrt{2}}\left[E + \frac{4uw\Omega_{c}^{2}e^{i(\theta+\frac{\pi}{2})}}{\sqrt{(2uw\Delta)^{2} + G^{2}_{\chi}}}\right]c_{1} .
\end{align}
It is apparent that there are two amplitudes contributing to the equations, i.e., one proportional to $E$ and the other proportional to $\Omega_{c}^{2}$ corresponding, respectively, to the two-photon and stepwise transitions from the ground state $\ket 1$ to the $\ket\xi$ and $\ket\zeta$ states. The difference between the amplitudes can be interpreted as a destructive interference. Thus, under the complete destructive interference, $\dot{c}_{\xi}=0$ and $\dot{c}_{\zeta}=0$. Hence, $c_{\xi}(t)$ and $c_{\zeta}(t)$ are constants, so if the amplitudes were zero at $t=0$, they remain zero for all times.

In practice, the method to determine the number of photons in a given system is to measure the one-time second-order correlation function. This function is directly related to photon-counting measurements, representing the joint  probability of two-photon detection. We calculate the second-order correlation function, $g^{(2)}(0) =\langle\hat{a}^{\dagger}\hat{a}^{\dagger}\hat{a}\hat{a}\rangle/\langle\hat{a}^{\dagger}\hat{a}\rangle^{2}$, of the cavity field in the steady state. The presence of only a single photon in the system is manifested by the photon antibunching effect, which signifies that at a given time, only a single photon is present in the system~\cite{cw76,km76,kd77}. The value $g^{(2)}(0)=0$ is the direct indication of the existence of only one photon in the system. As a check on the validity of the approximate analytical results, the calculations of $g^{(2)}(0)$ are carried out using a numerical method, i.e., the quantum optics toolbox for Matlab~\cite{ta99}, to solve the master equation exactly without the truncation of the Hilbert space of the system at $n=2$.

\begin{figure}[h]
\includegraphics[width=\columnwidth]{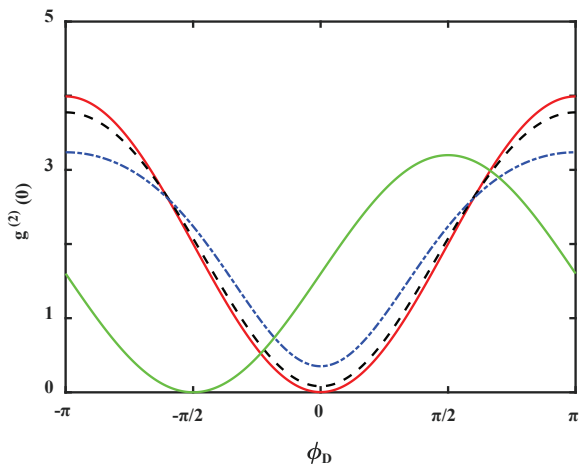}
\caption{(Color online) The second-order correlation function $g^{2}(0)$ of the cavity field plotted as a function of the phase $\phi_{D}$ of the DPO for $\chi =0$, $\Delta_{s}=\Delta =0$, $\gamma=\kappa$, $\Omega_{a}=\Omega_{c}$, $|E|=4\Omega_{c}^{2}/\kappa$, and several different values of $\Omega_{c}$: $\Omega_{c}=0.01\kappa$ (red solid line), $\Omega_{c}=0.05\kappa$ (black dashed line), and $\Omega_{c}=0.1\kappa$ (blue dash-dotted line). The solid green line shows $g^{(2)}(0)$ for a non-chiral situation of $\chi =1$ with $\Delta =\kappa$ and $\Omega_{c}=0.01\kappa$. }
\label{fig4}
\end{figure}

\begin{figure}[h]
\includegraphics[width=\columnwidth]{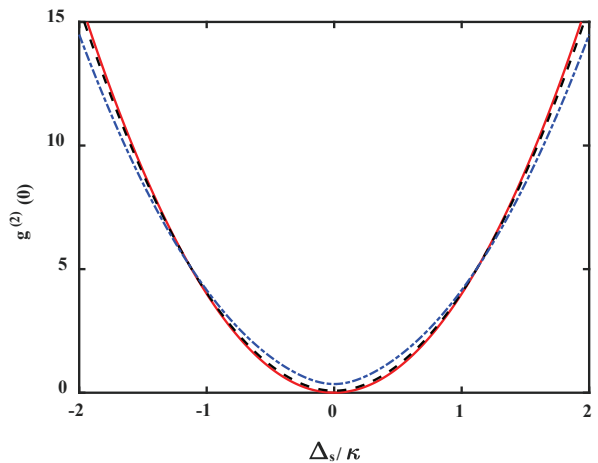}
\caption{(Color online) The second-order correlation function $g^{2}(0)$ of the cavity field plotted as a function of $\Delta_{s}$ for $\chi =0$, $\Delta =0$, $\gamma=\kappa$, $\Omega_{a}=\Omega_{c}$, $|E|=4\Omega_{c}^{2}/\kappa$ and different $\Omega_{c}$: $\Omega_{c}=0.01\kappa$ (red solid line), $\Omega_{c}=0.05\kappa$ (black dashed line), and $\Omega_{c}=0.1\kappa$ (blue dash-dotted line). }
\label{fig5}
\end{figure}

\begin{figure}[h]
\includegraphics[width=\columnwidth]{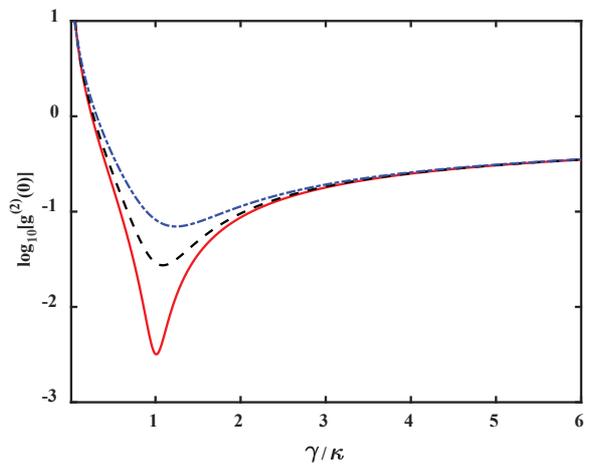}
\caption{(Color online) Variation of the second-order correlation function $g^{2}(0)$ with the damping rate $\gamma$ for $\chi =0$, $\Delta_{s}=\Delta =0$, $\Omega_{a}=\Omega_{c}$, $|E|=4\Omega_{c}^{2}/\kappa$ and various values of the Rabi frequency $\Omega_{c}$: $\Omega_{c}=0.01\kappa$ (red solid line), $\Omega_{c}=0.03\kappa$ (black dashed line), and $\Omega_{c}=0.05\kappa$ (blue dash-dotted line). }
\label{fig6}
\end{figure}

The second-order correlation function has been evaluated for certain combinations of the parameters determined by the conditions (\ref{c53}), and the results are shown in Figs.~\ref{fig4}$-$\ref{fig6}.
Let us first examine the variation of the correlation function with the phase of the DPO. Figure~\ref{fig4} shows $g^{(2)}(0)$ as a function of $\phi_{D}$  for several values of the Rabi frequency $\Omega_{c}$ and the values of the remaining parameters determined by the dark state conditions, given by Eq.~(\ref{c53}). For the perfect chiral case $(\chi=0)$, the complete photon antibunching, $g^{(2)}(0)=0$ is seen at $\phi_{D}=0$ and for a very weak driving field. Thus, there is a direct relationship between the dark state generation and the photon blockade effect.
Note that in the range of the Rabi frequency $\Omega_{c}\leq 0.05\kappa$, the change in the variation of $g^{(2)}(0)$ around zero is almost invisible.

Figure~\ref{fig4} also shows results for the non-chiral case $\chi=1$. It is seen that the complete photon antibunching is shifted to the phase $\phi_{D}= -\pi/2$. We point out that the positions of the complete photon antibunching predicted by the numerical results are properly accounted for by Eq.~(\ref{c53}). When $\Delta =0$, the phase angle $\theta=\pi/2$ and then the pure dark state is created for the phase $\phi_{D}=0$. On the other hand, when $\chi=1\, (G_{\chi}=0)$, the phase angle $\theta =0$, showing that the pure dark state is then created for the phase of the DPO shifted to $\phi_{D}=-\pi/2$. Thus it is evident that under a weak field excitation, the conditions for the production of the dark state obtained by solving the master equation in the truncated basis of $n\leq 2$ states agree perfectly with the numerical exact solution of the master equation. By moving away from the weak field approximation, differences between the two solutions become visible. Hence the approximation of neglecting the $n>2$ states is reasonable for a weak driving.

Figure~\ref{fig5} shows the variation of $g^{(2)}(0)$ with the detuning $\Delta_{s}$. The correlation function exhibits a dip centered about $\Delta_{s}=0$ and it is seen that for small Rabi frequencies of the driving field, $g^{(2)}(0)$, and therefore the joint probability of two-photon detection, vanishes. For the Rabi frequency in a range $\Omega_{c}\leq 0.05\kappa$, the curves are practically indistinguishable. Again, the perfect agreement between the exact numerical results and the prediction of the analytical results is evident.

The dark state condition (\ref{c53}) requires equal damping rates, $\kappa=\gamma$. Numerical results for this example are shown in Fig.~\ref{fig6}. It is seen that the correlation function dips very strongly at $\gamma=\kappa$. The dip is smoothed by an increased value of $\Omega_{c}$. Once again, we point out that $g^{(2)}(0)$ vanishes when the parameters satisfy the conditions for the production of the dark state.

It is notable that in Figs.~\ref{fig4}--\ref{fig6} the complete photon antibunching occurs at the values of the parameters corresponding to the creation of the dark state in the system. Therefore, the validity of the truncation of the Hilbert space at $n=2$ seems to be confirmed.
\begin{figure}[h]
\includegraphics[width=\columnwidth]{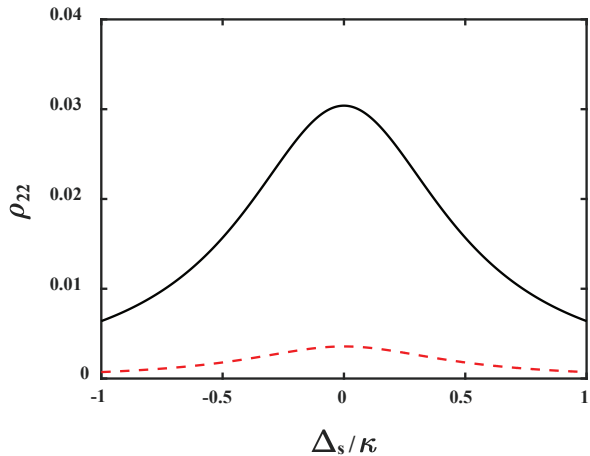}
\caption{(Color online) The steady-state population of the state $\ket 2\equiv \ket{g,1}$ plotted as a function of $\Delta_{s}$ for  $\gamma=\kappa$ for $\chi =0$, $\Delta =0$, $\Omega_{a}=\Omega_{c}$, $|E|=4\Omega_{c}^{2}/\kappa$ and various values of the Rabi frequency $\Omega_{c}$: $\Omega_{c}=0.03\kappa$ (red dashed line) and $\Omega_{c}=0.09\kappa$ (black solid line). }
\label{fig7}
\end{figure}

It is also worthwhile to calculate the population of the single-photon state $\ket 2\equiv \ket{g,1}$, representing the probability that the atom is in its ground state and there is a single photon present in the cavity. Note that the state $\ket 2$ contributes to the superposition state $\ket\phi$ involved in the dark state $\ket{\Psi_{2}}$. Figure~\ref{fig7} shows numerical results for the population $\rho_{22}$ versus $\Delta_{s}$, and the parameter values corresponding to the dark state are again used. The curves shown in the figure assume the optimal values of the Rabi frequency, $\Omega_{c}=0.03\kappa$ and $\Omega_{c}=0.09\kappa$, at which the complete photon antibunching is still present $(g^{(2)}(0)=0)$ when the Hilbert space is truncated at $n=1$ and $n=2$, respectively. We see that there is only a few percent probability of finding a photon in the cavity. However, we observe that in the weak driving limit corresponding to the $n=2$ truncation, the probability $\rho_{22}=0.03$ is about three times larger than that achieved in the conventional photon blockade schemes, where the probabilities $\rho_{22}<0.01$ were reported~\cite{ks14,fs16}.

\section{Summary}\label{sec5}

We have considered the dynamics of an open quantum system composed of a DPO cavity and a two-level atom coupled to a waveguide and driven by weak laser fields. The calculations have been carried out analytically in a Hilbert space truncated at the double-excitation states of the combined system. We have examined the conditions for the decoherence-free subspace and especially for the generation of a dark state and have found that there are two alternative ways  the system could decay to a pure dark state. One way to produce the pure state is to chirally couple the cavity and the atom to the waveguide. An alternative way is to drive the cavity and the atom with unequal detunings. The produced state is a coherent superposition of only the single-excitation and ground states of the combined system. In the case where the pure state is prepared by driving the systems with unequal detunings, the phase of the field pumping the DPO material must be shifted relative to the phase of the fields driving the cavity mode and the atom. However, in the case where the pure state is prepared by the chiral coupling of the systems to the waveguide, the phases of the pumping and driving fields should be equal.
In addition, we have found that the conditions for the production of the dark state coincide with the conditions for the photon blockade effect.
Finally, the analytic results for the conditions to produce the dark state were compared with the results obtained by the direct numerical integration of the master equation.
The perfect agreement between the results have been found in the limit of weak excitations.

This system provides a possible method for preparing a composite system in a pure dark state, which is a coherent superposition of the ground and excited single-photon state, by suitably choosing the detunings and Rabi frequencies of the driving fields.

\section*{Acknowledgements}

This work is supported by the National Natural Science Foundation of China (Grant No. 11474119).

\appendix

\section{Derivation of the master equation~(\ref{c2})}\label{App1}

In this appendix we give details of the derivation of the master equation for the reduced density operator of a system (atom plus cavity mode) interacting with the quantized one-dimensional
multimode field of the waveguide playing a role of a reservoir. We use the traditional technique based on the Born-Markov approximation~\cite{lh70,lo73}.

In the electric-dipole approximation, the Hamiltonian of the system interacting with the reservoir can be written in the following form $(\hbar\equiv 1)$:
\begin{align}
\hat{H} = \hat{H}_{sys}+\hat{H}_{res}+\hat{H}_{int} ,\label{A1}
 \end{align}
where
\begin{align}
\hat{H}_{sys} &= \omega_{c}\hat{a}^{\dagger}\hat{a} + \omega_{a}\hat{\sigma}_{+}\hat{\sigma}_{-} + \frac{1}{2}i\left(E^{\ast}\hat{a}^{2}e^{-i\omega_{p}t} - {\rm H.c.}\right) \nonumber\\
&+ i\left(\Omega_{c} \hat{a}e^{-i\omega_{L}t} +\Omega_{a} \hat{\sigma}_{-}e^{-i\omega_{L}t} - {\rm H.c.}\right) \label{A2}
\end{align}
is the Hamiltonian of the cavity mode and the atom driven by a laser field of frequency $\omega_{L}$, and a nonlinear medium pumped with a field of frequency $\omega_{p}$,
\begin{align}
\hat{H}_{res} = \sum_{\lambda=L,R}\int d\omega_{k_{\lambda}}\, \omega_{k_{\lambda}} \hat{c}^{\dag}_{k_{\lambda}}\hat{c}_{k_{\lambda}} \label{A3}
\end{align}
is the Hamiltonian of the reservoir, and
\begin{align}
\hat{H}_{int} =& -i\sum_{\lambda = L,R} \int d\omega_{k_{\lambda}} \left\{ \left[g_{c}(\omega_{k_{\lambda}})\hat{a}_{x}e^{ik_{\lambda} x_{c}}\right.\right. \nonumber\\
&\left.\left. +\, g_{a}(\omega_{k_{\lambda}})\hat{\sigma}_{x}e^{ik_{\lambda} x_{a}}\right]\hat{c}_{k_{\lambda}} -{\rm H.c.}\right\} \label{A4}
\end{align}
is the interaction of the cavity mode and the atom with the reservoir field.
Here, $\hat{\sigma}_{+}$ and $\hat{\sigma}_{-}$ are dipole raising and lowering operators for the atom, $\hat{a}^{\dagger}$ and $\hat{a}$ are creation and annihilation operators for the cavity mode, and 
$\hat{c}_{k_{\lambda}}$ are the bosonic annihilation operators of the reservoir mode $k_{\lambda}$ of frequency $\omega_{k_{\lambda}}$ propagating to the right $(\lambda=R)$ and to the left $(\lambda=L)$ directions.
The coefficients $g_{c}(\omega_{k_{\lambda}})$ and $g_{a}(\omega_{k_{\lambda}})$ describe the coupling of the cavity mode and the atom, respectively, with the waveguide field;
$\hat{a}_{x} = \hat{a} +\hat{a}^{\dag}$, $\hat{\sigma}_{x} = \hat{\sigma}_{-} +\hat{\sigma}_{+}$, and $k_{\lambda} =\omega_{k_{\lambda}}/\upsilon_{\lambda}$, in which $\upsilon_{\lambda}$ is the
group velocity of the modes.

The reservoir operators satisfy the following Heisenberg equation of motion
\begin{align}
\dot{\hat{c}}_{k_{\lambda}} =& -i\omega_{k_{\lambda}}\hat{c}_{k_{\lambda}} + g_{c}(\omega_{k_{\lambda}})\hat{a}_{x} e^{-ik_{\lambda} x_{c}} \nonumber\\
& +\, g_{a}(\omega_{k_{\lambda}})\hat{\sigma}_{x} e^{-ik_{\lambda} x_{a}} ,\label{A5}
\end{align}
which may be integrated formally and the solution is
\begin{align}
\hat{c}_{k_{\lambda}}(t) &= \hat{c}_{k_{\lambda}}(0)e^{-i\omega_{k_{\lambda}}t} \nonumber\\
&+g_{c}(\omega_{k_{\lambda}})e^{-ik_{\lambda} x_{c}}\!\!\int_{0}^{t} \!ds\, \hat{a}_{x}(s) e^{-i\omega_{k_{\lambda}}(t-s)} \nonumber\\
& +\, g_{a}(\omega_{k_{\lambda}})e^{-ik_{\lambda} x_{a}}\!\!\int_{0}^{t}\!ds\, \hat{\sigma}_{x}(s)e^{-i\omega_{k_{\lambda}}(t-s)} .\label{A6}
\end{align}
The time integrals appearing in Eq.~(\ref{A6}) can be evaluated with the following two approximations, the Born approximation assuming weak coupling of the systems to the reservoir field, and the Markov approximation assuming that the time $t\gg \tau_{r}$, where $\tau_{r}$ is the correlation time of the reservoir. In this case, we may write the operators $\hat{a}_{x}(s)$ and $\hat{\sigma}_{x}(s)$ in terms of slowly and quickly varying parts,
\begin{align}
\hat{a}_{x}(s) &= \hat{a}(t)e^{i\omega_{L}(t-s)} + \hat{a}^{\dag}(t)e^{-i\omega_{L}(t-s)} ,\nonumber\\
\hat{\sigma}_{x}(s) &= \hat{\sigma}_{-}(t)e^{i\omega_{L}(t-s)} + \hat{\sigma}_{+}(t)e^{-i\omega_{L}(t-s)} .
\end{align}
The slowly varying parts of the operators are evaluated at time $t$, since over the short correlation time of the reservoir, the slowly varying parts of $\hat{a}(s)$ and $\hat{\sigma}_{-}(s)$ would hardly change from $\hat{a}(t)$ and $\hat{\sigma}_{-}(t)$. Thus, the time integrals appearing in Eq.~(\ref{A6}) can be evaluated to give
\begin{align}
\int_{0}^{t}\!ds\, e^{i(\omega_{k_{\lambda}}\pm\omega_{L})(t-s)} = \pi\delta(\omega_{k_{\lambda}}\!\pm\!\omega_{L}) + i\frac{P}{\omega_{k_{\lambda}}\!\pm\!\omega_{L}}  \equiv J_{\pm} ,\label{A7}
\end{align}
in which $P$ stands for the principal value. Hence, we can approximate Eq.~(\ref{A6}) by
\begin{align}
\hat{c}_{k_{\lambda}}(t) &= \hat{c}_{0}(t) +g_{c}(\omega_{k_{\lambda}})e^{-i\omega_{k_{\lambda}} \tau_{c}}\left[\hat{a}(t)J_{-}^{\ast} +\hat{a}^{\dag}(t)J_{+}^{\ast}\right] \nonumber\\
&+\, g_{a}(\omega_{k_{\lambda}})e^{-i\omega_{k_{\lambda}} \tau_{a}}\left[\hat{\sigma}_{-}(t)J_{-}^{\ast} + \hat{\sigma}_{+}(t)J_{+}^{\ast}\right] ,\label{A8}
\end{align}
where $\hat{c}_{0}(t) = \hat{c}_{k_{\lambda}}(0)\exp(-i\Delta_{k_{\lambda}}t)$.

Suppose that $\hat{\mathcal{O}}(t)$ is an arbitrary combination of atomic or cavity mode operators. We make the unitary transformation
\begin{align}
\tilde{\mathcal{O}}(t) =e^{i\hat{H}_{0}t}\hat{\mathcal{O}}(t)e^{-i\hat{H}_{0}t} ,
\end{align}
where
\begin{align}
\hat{H}_{0} = \omega_{L}\hat{a}^{\dagger}\hat{a} + \omega_{L}\hat{\sigma}_{+}\hat{\sigma}_{-} ,
\end{align}
and find the Heisenberg equation of motion for $\tilde{\mathcal{O}}(t)$:
\begin{align}
\dot{\tilde{\mathcal{O}}}(t) =& -i\left[\tilde{\mathcal{O}}(t),\tilde{H}_{sys}\right] \nonumber\\
&+\!\sum_{\lambda=L,R}\int \!\!d\omega_{k_{\lambda}}g_{c}(\omega_{k_{\lambda}})\!\left\{\hat{c}^{\dag}_{k_{\lambda}}\!(t)\!\!\left[\tilde{\mathcal{O}}(t),\hat{a}_{x}(t)\right]\!e^{-ik_{\lambda} x_{c}}\right. \nonumber\\
&\left. + \left[\tilde{\mathcal{O}}(t),\hat{a}_{x}(t)\right]\hat{c}_{k_{\lambda}}(t)e^{ik_{\lambda} x_{c}}\right\} \nonumber\\
&+\!\sum_{\lambda=L,R}\int\!\!d\omega_{k_{\lambda}}g_{a}(\omega_{k_{\lambda}})\!\left\{\hat{c}^{\dag}_{k_{\lambda}}\!(t)\!\!\left[\tilde{\mathcal{O}}(t),\hat{\sigma}_{x}(t)\right]\!e^{-ik_{\lambda} x_{a}}\right. \nonumber\\
&\left. +\left[\tilde{\mathcal{O}}(t),\hat{\sigma}_{x}(t)\right]\hat{c}_{k_{\lambda}}(t)e^{ik_{\lambda} x_{a}}\right\}  ,\label{A9}
\end{align}
where
\begin{align}
\tilde{H}_{sys} &= \Delta_{c}\hat{a}^{\dagger}\hat{a} + \Delta_{a}\hat{\sigma}_{+}\hat{\sigma}_{-} + \frac{1}{2}i\left(E^{\ast}\hat{a}^{2}- {\rm H.c.}\right) \nonumber\\
&+ i\left(\Omega_{c} \hat{a} +\Omega_{a} \hat{\sigma}_{-} - {\rm H.c.}\right) .
\end{align}

Substituting Eq.~(\ref{A8}) into Eq.~(\ref{A9}) and making the rotating-wave approximation in which we ignore all terms oscillating with frequency $2\omega_{L}$, we obtain
\begin{align}
\dot{\hat{\mathcal{O}}} =& -i\left[\hat{\mathcal{O}},\hat{H}^{\prime}_{sys}\right] -i\Omega_{ac}\left[\hat{\mathcal{O}}, \hat{\sigma}_{+}\hat{a} +\hat{a}^{\dag}\hat{\sigma}_{-}\right] \nonumber\\
&+ \hat{E}_{0}^{\dag}(x_{c},t)\left[\hat{\mathcal{O}},\hat{a}\right] +\hat{E}_{0}^{\dag}(x_{a},t)\left[\hat{\mathcal{O}},\hat{\sigma}_{-}\right]\nonumber\\
&+\hat{E}_{0}(x_{c},t)\left[\hat{\mathcal{O}},\hat{a}^{\dag}\right] +\hat{E}_{0}(x_{a},t)\left[\hat{\mathcal{O}},\hat{\sigma}_{+}\right] \nonumber\\
&+ \frac{1}{2}\sum_{\lambda=L,R}\kappa_{\lambda}\left\{\hat{a}^{\dag}\left[\hat{\mathcal{O}},\hat{a}\right] -\left[\hat{\mathcal{O}},\hat{a}^{\dag}\right]\hat{a}\right\} \nonumber\\
&+ \frac{1}{2}\sum_{\lambda=L,R}\gamma_{\lambda}\left\{\hat{\sigma}_{+}\left[\hat{\mathcal{O}},\hat{\sigma}_{-}\right] -\left[\hat{\mathcal{O}},\hat{\sigma}_{+}\right]\hat{\sigma}_{-}\right\} \nonumber\\
&+ \frac{1}{2}\gamma_{ac}^{+}\left\{\hat{\sigma}_{+}\left[\hat{\mathcal{O}},\hat{a}\right] -\left[\hat{\mathcal{O}},\hat{\sigma}_{+}\right]\hat{a}\right\} \nonumber\\
&+ \frac{1}{2}\gamma^{-}_{ac}\left\{\hat{a}^{\dag}\left[\hat{\mathcal{O}},\hat{\sigma}_{-}\right] -\left[\hat{\mathcal{O}},\hat{a}^{\dag}\right]\hat{\sigma}_{-}\right\}  ,\label{A11}
\end{align}
where
\begin{align}
\hat{H}^{\prime}_{sys} &= \left(\Delta_{c}-\Delta_{c}^{-}\right)\hat{a}^{\dagger}\hat{a} -\Delta_{c}^{+}\hat{a}\hat{a}^{\dag} \nonumber\\
&+ \left(\Delta_{a}-\Delta_{a}^{-}\right)\hat{\sigma}_{+}\hat{\sigma}_{-} -\Delta_{a}^{+}\hat{\sigma}_{-}\hat{\sigma}_{+} \nonumber\\
&+ \frac{1}{2}i\left(E^{\ast}\hat{a}^{2}- {\rm H.c.}\right) + i\left(\Omega_{c} \hat{a} +\Omega_{a} \hat{\sigma}_{-} - {\rm H.c.}\right) ,\nonumber\\
\hat{E}_{0}(x_{c},t) &= \sum_{\lambda=L,R}\int d\omega_{k_{\lambda}}g_{c}(\omega_{k_{\lambda}})\hat{c}_{0}(t)e^{ik_{\lambda} x_{c}} ,\nonumber\\
\hat{E}_{0}(x_{a},t) &= \sum_{\lambda=L,R}\int d\omega_{k_{\lambda}}g_{a}(\omega_{k_{\lambda}})\hat{c}_{0}(t)e^{ik_{\lambda} x_{a}} ,\label{A12}
\end{align}
and
\begin{align}
\kappa_{\lambda} &=  2\pi\int\!d\omega_{k_{\lambda}}g^{2}_{c}(\omega_{k_{\lambda}})\delta(\omega_{k_{\lambda}}-\omega_{L}) ,\nonumber\\
\gamma_{\lambda} &=  2\pi \int\!d\omega_{k_{\lambda}}g^{2}_{a}(\omega_{k_{\lambda}})\delta(\omega_{k_{\lambda}} - \omega_{L}) ,\nonumber\\
\gamma^{\pm}_{ac} &= 2\pi\sum_{\lambda=L,R}\int d\omega_{k_{\lambda}}g_{a}(\omega_{k_{\lambda}})g_{c}(\omega_{k_{\lambda}}) \nonumber\\
&\times e^{\pm ik_{\lambda} x_{ac}}\delta(\omega_{k_{\lambda}}-\omega_{L}) ,\nonumber\\
\Omega_{ac} &= -\sum_{\lambda=L,R} P \int d\omega_{k_{\lambda}}g_{a}(\omega_{k_{\lambda}})g_{c}(\omega_{k_{\lambda}}) \nonumber\\
&\times \left(\frac{e^{ik_{\lambda}x_{ac}}}{\omega_{k_{\lambda}} -\omega_{L}} + \frac{e^{-ik_{\lambda} x_{ac}}}{\omega_{k_{\lambda}} +\omega_{L}}\right) ,\nonumber\\
\Delta_{a(c)}^{\pm} &= \sum_{\lambda=L,R} P \int d\omega_{k_{\lambda}} \frac{g^{2}_{a(c)}(\omega_{k_{\lambda}})}{\omega_{k_{\lambda}} \pm\omega_{L}} ,\label{A13}
\end{align}
with $x_{ac}=x_{a}-x_{c}$ and $\omega_{p} = 2\omega_{L}$.
The parameters $\kappa$ and $\gamma$, which appear in Eq.~(\ref{A11}), are the damping rates of the cavity mode and the atom, respectively. The parameters $\gamma^{\pm}_{ac}$ are decay constants which arise from the dissipative coupling of the cavity mode with the atom induced by the waveguide field. Moreover, Eq.~(\ref{A11}) also contains a distance dependent constant $\Omega_{ac}$ which arises from a coherent exchange of an excitation between the cavity mode and the atom induced by the waveguide field. The terms $\Delta_{a(c)}^{\pm}$ represent the Lamb shifts of the energy levels of the systems. There terms might be omitted as the Lamb shift is small or could be included in the detunings $\Delta_{c}$ and $\Delta_{a}$ by redefining the frequencies, $\omega_{c}\rightarrow \omega_{c}-(\Delta_{c}^{-}+\Delta_{c}^{+})$ and $\omega_{a}\rightarrow \omega_{a}-(\Delta_{a}^{-}-\Delta_{a}^{+})$.

Using the relations
\begin{align}
g_{a}(\omega_{k_{\lambda}}) = g_{a}\sqrt{\omega_{k_{\lambda}}} ,\quad  g_{c}(\omega_{k_{\lambda}}) = g_{c}\sqrt{\omega_{k_{\lambda}}} ,
\end{align}
where $g_{a}$ and $g_{c}$ are constants, we can evaluate the parameters and find
\begin{align}
\kappa_{\lambda} &=  2\pi g^{2}_{c}(\omega_{L_{\lambda}}) ,\quad
\gamma_{\lambda} =  2\pi g^{2}_{a}(\omega_{L_{\lambda}}) ,\nonumber\\
\gamma^{+}_{ac} = \gamma^{-}_{ac} &= 4\pi g_{a}(\omega_{L})g_{c}(\omega_{L})\cos(k_{0} x_{ac}) ,\nonumber\\
\Omega_{ac} &= 2\pi g_{a}(\omega_{L})g_{c}(\omega_{L})\sin(k_{0}x_{ac}) ,\label{A15}
\end{align}
in which $k_{0}\equiv k_{R}$, and $\omega_{L_{\lambda}}$ describes the waveguide mode of frequency $\omega_{L}$ propagating in the $\lambda$ direction. In the derivation of the expressions for $\Omega_{ac}$ and $\gamma^{\pm}_{ac}$, we have taken $k_{L}= -k_{R}$, which results from the fact that $\upsilon_{R}>0$ and $\upsilon_{L}<0$.

If we take the expectation value of both sides of Eq.~(\ref{A11}) over the initial state of the entire system, assuming that initially the reservoir field was in the vacuum state $\ket 0$,
and note that
\begin{align}
\langle\dot{\hat{\mathcal{O}}}\rangle = {\rm Tr}\{\dot{\hat{\mathcal{O}}}\rho\} ={\rm Tr}\{\hat{\mathcal{O}}\dot{\rho}\} ,
\end{align}
where $\rho$ is the density operator describing the properties of the "reduced" system, the cavity mode plus the atom, we readily find that Eq.~(\ref{A11}) transfers to
\begin{align}
\dot{\rho} =& -i\left[\hat{H}^{\prime}_{sys},\rho\right] +i\Omega_{ac}\left[\rho,\hat{\sigma}_{+}\hat{a} +\hat{a}^{\dag}\hat{\sigma}_{-}\right] \nonumber\\
&+\frac{1}{2}\sum_{\lambda=L,R}\kappa_{\lambda}\left(\left[\hat{a},\rho\hat{a}^{\dag}\right] -\left[\hat{a}^{\dag},\hat{a}\rho\right]\right) \nonumber\\
&+\frac{1}{2}\sum_{\lambda=L,R}\gamma_{\lambda}\left(\left[\hat{\sigma}_{-},\rho\hat{\sigma}_{+}\right] -\left[\hat{\sigma}_{+},\hat{\sigma}_{-}\rho\right]\right) \nonumber\\
&+\sqrt{\kappa_{L}\gamma_{L}}\cos(k_{0} x_{ac})\!\left(\left[\hat{a},\rho\hat{\sigma}_{+}\right] -\left[\hat{a}^{\dag},\hat{\sigma}_{-}\rho\right]\right) \nonumber\\
&+\sqrt{\kappa_{R}\gamma_{R}}\cos(k_{0} x_{ac})\!\left(\left[\hat{\sigma}_{-},\rho\hat{a}^{\dag}\right] -\left[\hat{\sigma}_{+},\hat{a}\rho\right]\right) ,\label{A17}
\end{align}
which is in the form of the master equation for the density operator of the system. In Eq.~(\ref{A17}), the dissipative cross-coupling terms have been divided into two parts corresponding to the excitation propagating to the right and to the left in accordance with the properties of a cascaded quantum system~\cite{wg93,hc93,kc94,gp94,nc05,gz04} where only the excitation of the cavity mode propagating to the right can couple to the atom and vice versa, i.e., only the excitation of the atom propagating to the left can couple to the cavity mode.

\section{Representations of the collective operators}\label{App2}

In this appendix, we give the explicit forms of the superposition operators in terms of the projection operators between the product states~(\ref{c8a}).

Applying  the completeness relations
\begin{align}
\sum_{n=0}^{\infty}\ket n\bra n =1 \quad {\rm and}\quad \ket g\bra g +\ket e \bra e =1 ,\label{B1}
\end{align}
to the operators $\hat{J}^{\dagger} = u\hat{a}^{\dagger} +w\hat{\sigma}_{+}$ and $\hat{B}^{\dagger} = w\hat{a}^{\dagger} - u\hat{\sigma}_{+}$, we get
\begin{align}
\hat{J}^{\dagger} &=  u\!\left(\ket{2}\bra{1} +\ket{5}\bra{4} +\sqrt{2}\ket{3}\bra{2}\right) + w\!\left(\ket{4}\bra{1} +\ket{5}\bra{2}\right) ,\nonumber\\
\hat{B}^{\dagger} &=  w\!\left(\ket{2}\bra{1} +\ket{5}\bra{4} +\sqrt{2}\ket{3}\bra{2}\right) -u\left(\ket{4}\bra{1} +\ket{5}\bra{2}\right) ,\label{B2}
\end{align}
Then, using the results of Eq.~(\ref{B2}), we get
\begin{align}
\hat{J}^{\dagger}\hat{J} &= u^{2}\hat{a}^{\dagger}\hat{a} +w^{2}\hat{\sigma}_{+}\hat{\sigma}_{-} +uw\!\left( \hat{a}^{\dagger}\hat{\sigma}_{-} +\hat{\sigma}_{+} \hat{a}\right) ,\nonumber\\
\hat{B}^{\dagger}\hat{B} &= w^{2}\hat{a}^{\dagger}\hat{a} +u^{2}\hat{\sigma}_{+}\hat{\sigma}_{-} -uw\!\left( \hat{a}^{\dagger}\hat{\sigma}_{-} +\hat{\sigma}_{+}\hat{a}\right) ,\nonumber\\
\hat{J}^{\dagger}\hat{B} &= w^{2}\hat{\sigma}_{+}\hat{a} -u^{2}\hat{a}^{\dagger}\hat{\sigma}_{-} +uw\!\left( \hat{a}^{\dagger}\hat{a} -\hat{\sigma}_{+} \hat{\sigma}_{-}\right) ,\nonumber\\
\hat{B}^{\dagger}\hat{J} &= w^{2}\hat{a}^{\dagger}\hat{\sigma}_{-}\!-\!u^{2}\hat{\sigma}_{+}\hat{a} + uw\!\left( \hat{a}^{\dagger}\hat{a} -\hat{\sigma}_{+} \hat{\sigma}_{-}\right) ,\label{B3}
\end{align}
with
\begin{align}
\hat{a}^{\dagger}\hat{a} &= \ket{2}\bra{2} +\ket{5}\bra{5} + 2 \ket{3}\bra{3} ,\nonumber\\
\hat{\sigma}_{+}\hat{\sigma}_{-} &= \ket{4}\bra{4} +\ket{5}\bra{5} ,\nonumber\\
\hat{\sigma}_{+}\hat{a} &= \ket{4}\bra{2} +\sqrt{2}\ket{5}\bra{3} ,\nonumber\\
\hat{a}^{\dagger}\hat{\sigma}_{-} &= \ket{2}\bra{4} +\sqrt{2}\ket{3}\bra{5} .\label{B4}
\end{align}

\section{Equations of motion for the density matrix elements}\label{App3}

In this appendix, we present explicitly the complete set of the equations of motion for the density matrix elements in the basis spanned by the superposition states $\{\ket 1,\ket\psi,\ket\phi,\ket\xi,\ket\zeta\}$. The equations have been derived from the master equation (\ref{c11}) using the explicit forms of the superposition operators listed in Appendix~\ref{App1}. We have introduced density matrix elements with respect to the superposition states, denoting $\bra\psi\rho\ket\phi$ by $\rho_{\psi\phi}$, etc., and have written separately the equations of motion for the populations and coherences. Thus the equations for the populations of the states are
\begin{align}
\dot{\rho}_{\phi\phi} =& -G_{\chi}\left(\rho_{\psi\phi}+\rho_{\phi\psi}\right) -2iuw\Delta\left(\rho_{\psi\phi}-\rho_{\phi\psi}\right) \nonumber\\
                       &+\Gamma_{\xi\phi}\rho_{\xi\xi} +\Gamma_{\zeta\phi}\rho_{\zeta\zeta} +\sqrt{\Gamma_{\xi\phi}\Gamma_{\zeta\phi}}(\rho_{\xi\zeta}+\rho_{\zeta\xi})\nonumber\\
                       &+\Omega_{\phi\xi}(\rho_{\xi\phi}\!+\!\rho_{\phi\xi}) +\Omega_{\phi\zeta}(\rho_{\zeta\phi}\!+\!\rho_{\phi\zeta}) \nonumber\\
                       &-\Omega_{\phi}(\rho_{1\phi} + \rho_{\phi1}) ,\nonumber\\
\dot{\rho}_{\psi\psi} =& -\Gamma_{\chi}\rho_{\psi\psi} + G_{\chi}\left(\rho_{\psi\phi}+\rho_{\phi\psi}\right)\!+\!2iuw\Delta\left(\rho_{\psi\phi}-\rho_{\phi\psi}\right) \nonumber\\
&+\Gamma_{\xi\psi}\rho_{\xi\xi} +\Gamma_{\zeta\phi}\rho_{\zeta\zeta} +\sqrt{\Gamma_{\xi\psi}\Gamma_{\zeta\psi}}(\rho_{\xi\zeta}+\rho_{\zeta\xi}) \nonumber\\
&+\Omega_{\psi\zeta}(\rho_{\psi\zeta} + \rho_{\zeta\psi}) +\Omega_{\psi\xi}(\rho_{\psi\xi}\!+\!\rho_{\xi\psi}) \nonumber\\
&-\Omega_{\psi}(\rho_{1\psi} + \rho_{\psi1}) ,\nonumber\\
\dot{\rho}_{\xi\xi} =& -\left(\Gamma_{\xi\phi} + \Gamma_{\xi\psi}\right)\rho_{\xi\xi} + 2i\alpha\beta\Delta (\rho_{\xi\zeta} -\rho_{\zeta\xi}) \nonumber\\
                      &-\frac{1}{2}\left(\sqrt{\Gamma_{\xi\phi}\Gamma_{\zeta\phi}}\!+\!\sqrt{\Gamma_{\xi\psi}\Gamma_{\zeta\psi}}\!-\!2\sqrt{2}G_{\chi}\right)\left(\rho_{\xi\zeta}+\rho_{\zeta\xi}\right) \nonumber\\
                      & -\Omega_{\psi\xi}(\rho_{\xi\psi}+\rho_{\psi\xi}) -\Omega_{\phi\xi}(\rho_{\xi\phi}+\rho_{\phi\xi}) \nonumber\\
                      &-\alpha(\tilde{E}\rho_{1\xi}+\tilde{E}^{\ast}\rho_{\xi1}) ,\nonumber\\
\dot{\rho}_{\zeta\zeta} =& -\left(\Gamma_{\zeta\phi} + \Gamma_{\zeta\psi}\right)\rho_{\zeta\zeta} - 2i\alpha\beta\Delta (\rho_{\xi\zeta} -\rho_{\zeta\xi}) \nonumber\\
&-\frac{1}{2}\left(\sqrt{\Gamma_{\xi\phi}\Gamma_{\zeta\phi}}\!+\!\sqrt{\Gamma_{\xi\psi}\Gamma_{\zeta\psi}}\!+\!2\sqrt{2}G_{\chi}\right)\!\left(\rho_{\xi\zeta}+\rho_{\zeta\xi}\right) \nonumber\\
                       &-\Omega_{\psi\zeta}(\rho_{\zeta\psi}+\rho_{\psi\zeta}) -\Omega_{\phi\zeta}(\rho_{\zeta\phi}+\rho_{\phi\zeta}) \nonumber\\
                       &-\beta( \tilde{E}\rho_{1\zeta}+\tilde{E}^{\ast}\rho_{\zeta1}) .\label{C1}
\end{align}
The equations of motion for the coherences between the ground and excited states are
\begin{align}
\dot{\rho}_{1\psi} =& -\left\{\frac{1}{2}\Gamma_{\chi} -i\!\left[\Delta_{s}+(u^{2}\!-\!w^{2})\Delta\right]\right\}\rho_{1\psi} \nonumber\\
&+\sqrt{\Gamma_{\chi}\Gamma_{\zeta\psi}}\rho_{\psi\zeta} + \sqrt{\Gamma_{\chi}\Gamma_{\xi\psi}}\rho_{\psi\xi} \nonumber\\
&+(G_{\chi}+2iuw\Delta)\rho_{1\phi} +\tilde{E^{\ast}}(\beta\rho_{\zeta\psi}+\alpha\rho_{\xi\psi}) \nonumber\\
&+\Omega_{\psi}(\rho_{\psi\psi} - \rho_{11}) +\Omega_{\phi}\rho_{\phi\psi} +\Omega_{\psi\zeta}\rho_{1\zeta} +\Omega_{\psi\xi}\rho_{1\xi} ,\nonumber\\
\dot{\rho}_{1\phi} =&\  i\!\left[\Delta_{s}\!-\!(u^{2}\!-\!w^{2})\Delta\right]\!\rho_{1\phi}\!+\!\sqrt{\Gamma_{\chi}\Gamma_{\zeta\phi}}\rho_{\psi\zeta}\!+\!\sqrt{\Gamma_{\chi}\Gamma_{\xi\phi}}\rho_{\psi\xi} \nonumber\\
&-(G_{\chi}-2iuw\Delta)\rho_{1\psi} +\tilde{E^{\ast}}(\beta\rho_{\zeta\phi}+\alpha\rho_{\xi\phi}) \nonumber\\
                    &+\Omega_{\phi}(\rho_{\phi\phi} -\rho_{11}) +\Omega_{\psi}\rho_{\psi\phi} +\Omega_{\phi\xi}\rho_{1\xi} +\Omega_{\phi\zeta}\rho_{1\zeta} ,\nonumber\\
\dot{\rho}_{1\xi} =& -\left[\frac{1}{2}(\Gamma_{\xi\psi} +\Gamma_{\xi\phi}) -2i\left(\Delta_{s}+\alpha^{2}\Delta\right)\right]\rho_{1\xi} \nonumber\\
&-\left[\frac{1}{2}\!\left(\sqrt{\Gamma_{\xi\phi}\Gamma_{\zeta\phi}}\!+\!\sqrt{\Gamma_{\xi\psi}\Gamma_{\zeta\psi}}\!-\!2\sqrt{2}G_{\chi}\!\right)\!-\!2i\alpha\beta\Delta\right]\!\rho_{1\zeta} \nonumber\\
&+\tilde{E}^{\ast}\left[\beta\rho_{\zeta\xi}+\alpha\left(\rho_{\xi\xi}-\rho_{11}\right)\right]+\Omega_{\psi}\rho_{\psi\xi}+\Omega_{\phi}\rho_{\phi\xi}\nonumber\\
                     &-\Omega_{\phi\xi}\rho_{1\phi} -\Omega_{\psi\xi}\rho_{1\psi} ,\nonumber\\
\dot{\rho}_{1\zeta} =& -\left[\frac{1}{2}(\Gamma_{\zeta\psi} +\Gamma_{\zeta\phi}) -2i\left(\Delta_{s}+\beta^{2}\Delta\right)\right]\rho_{1\zeta} \nonumber\\
&-\left[\frac{1}{2}\!\left(\!\sqrt{\Gamma_{\xi\phi}\Gamma_{\zeta\phi}}\!+\!\sqrt{\Gamma_{\xi\psi}\Gamma_{\zeta\psi}}\!+\!2\sqrt{2}G_{\chi}\!\right)\!-\!2i\alpha\beta\Delta\right]\!\rho_{1\xi} \nonumber\\
&+\tilde{E^{\ast}}(\alpha\rho_{\xi\zeta}+\beta\rho_{\zeta\zeta}-\beta\rho_{11})+\Omega_{\psi}\rho_{\psi\zeta}+\Omega_{\phi}\rho_{\phi\zeta}\nonumber\\
                    &-\Omega_{\psi\zeta}\rho_{1\psi} -\Omega_{\phi\zeta}\rho_{1\phi} ,\label{C2}
\end{align}
and the equations of motion for the coherences between the excited states are
\begin{align}
\dot{\rho}_{\psi\phi} =& -\left[\frac{1}{2}\Gamma_{\chi} + 2i(u^{2}-w^{2})\Delta\right]\rho_{\psi\phi} +\sqrt{\Gamma_{\xi\psi}\Gamma_{\xi\phi}}\rho_{\xi\xi} \nonumber\\
&+\sqrt{\Gamma_{\zeta\psi}\Gamma_{\zeta\phi}}\rho_{\zeta\zeta} +\sqrt{\Gamma_{\zeta\psi}\Gamma_{\xi\phi}}\rho_{\zeta\xi} +\sqrt{\Gamma_{\xi\psi}\Gamma_{\zeta\phi}}\rho_{\xi\zeta} \nonumber\\
&+(G_{\chi}-2iuw\Delta)(\rho_{\phi\phi}-\rho_{\psi\psi}) -\Omega_{\psi}\rho_{1\phi}-\Omega_{\phi}\rho_{\psi1} \nonumber\\
&+\Omega_{\phi\xi}\rho_{\psi\xi} +\Omega_{\psi\xi}\rho_{\xi\phi} +\Omega_{\psi\zeta}\rho_{\zeta\phi} +\Omega_{\phi\zeta}\rho_{\psi\zeta} ,\nonumber\\
\dot{\rho}_{\psi\xi} =& -\frac{1}{2}\!\left\{\Gamma_{\chi}\!+\!\Gamma_{\xi\psi}\!+\!\Gamma_{\xi\phi}\!-\!2i\!\left[\Delta_{s}\!+\!(2\alpha^{2}\!-\!u^{2}\!+\!w^{2})\Delta\right]\!\right\}\!\rho_{\psi\xi} \nonumber\\
&-\left[\frac{1}{2}\!\left(\!\sqrt{\Gamma_{\xi\phi}\Gamma_{\zeta\phi}}\!+\!\sqrt{\Gamma_{\xi\psi}\Gamma_{\zeta\psi}}\!-\!2\sqrt{2}G_{\chi}\!\right)\!-\!2i\alpha\beta\Delta\right]\!\rho_{\psi\zeta} \nonumber\\
&-\alpha\tilde{E}^{\ast}\rho_{\psi1}+(G_{\chi}-2iuw\Delta)\rho_{\phi\xi} \nonumber\\
&-\Omega_{\psi}\rho_{1\xi}+\Omega_{\psi\xi}(\rho_{\xi\xi}-\rho_{\psi\psi}) +\Omega_{\psi\zeta}\rho_{\zeta\xi} -\Omega_{\phi\xi}\rho_{\psi\phi}  ,\nonumber\\
\dot{\rho}_{\psi\zeta} =& -\frac{1}{2}\!\left\{\Gamma_{\zeta\psi}\!+\!\Gamma_{\zeta\phi} -2i\!\left[\Delta_{s}\!+\!(2\beta^{2}\!-\!u^{2}\!+\!w^{2})\Delta\right]\right\}\rho_{\psi\zeta} \nonumber\\
&-\left[\frac{1}{2}\left(\!\sqrt{\Gamma_{\xi\phi}\Gamma_{\zeta\phi}}\!+\!\sqrt{\Gamma_{\xi\psi}\Gamma_{\zeta\psi}}\!+\!2\sqrt{2}G_{\chi}\right)\!-\!2i\alpha\beta\Delta\right]\!\rho_{\psi\xi} \nonumber\\
&-\beta\tilde{E}^{\ast}\rho_{\psi1}+(G_{\chi}-2iuw\Delta)\rho_{\phi\zeta} \nonumber\\
&-\Omega_{\psi}\rho_{1\zeta} -\Omega_{\psi\zeta}(\rho_{\psi\psi}-\rho_{\zeta\zeta}) -\Omega_{\phi\zeta}\rho_{\psi\phi} +\Omega_{\psi\xi} \rho_{\xi\zeta} ,\nonumber
\end{align}
\begin{align}
\dot{\rho}_{\phi\xi} =& -\frac{1}{2}\!\left\{\Gamma_{\xi\psi}\!+\!\Gamma_{\xi\phi} -2i\!\left[\Delta_{s}\!+\!(2\alpha^{2}\!+\!u^{2}\!-\!w^{2})\Delta\right]\right\}\rho_{\phi\xi} \nonumber\\
&-\left[\frac{1}{2}\left(\!\sqrt{\Gamma_{\xi\phi}\Gamma_{\zeta\phi}}\!+\!\sqrt{\Gamma_{\xi\psi}\Gamma_{\zeta\psi}}\!-\!2\sqrt{2}G_{\chi}\right)\!-\!2i\alpha\beta\Delta\right]\!\rho_{\phi\zeta} \nonumber\\
&-\alpha\tilde{E}^{\ast}\rho_{\phi1}-(G_{\chi}+2iuw\Delta)\rho_{\psi\xi} \nonumber\\
&-\Omega_{\phi}\rho_{1\xi}+\Omega_{\phi\xi}(\rho_{\xi\xi}-\rho_{\phi\phi}) +\Omega_{\phi\zeta}\rho_{\zeta\xi} -\Omega_{\psi\xi}\rho_{\phi\psi}  ,\nonumber\\
\dot{\rho}_{\phi\zeta} =& -\frac{1}{2}\!\left\{\Gamma_{\zeta\psi}\!+\!\Gamma_{\zeta\phi} -2i\!\left[\Delta_{s}\!+\!(2\beta^{2}\!+\!u^{2}\!-\!w^{2})\Delta\right]\right\}\rho_{\phi\zeta} \nonumber\\
&-\left[\frac{1}{2}\left(\!\sqrt{\Gamma_{\xi\phi}\Gamma_{\zeta\phi}}\!+\!\sqrt{\Gamma_{\xi\psi}\Gamma_{\zeta\psi}}\!+\!2\sqrt{2}G_{\chi}\right)\!-\!2i\alpha\beta\Delta\right]\!\rho_{\phi\xi} \nonumber\\
&-\beta\tilde{E}^{\ast}\rho_{\phi1}-(G_{\chi}+2iuw\Delta)\rho_{\psi\zeta} \nonumber\\
&-\Omega_{\phi}\rho_{1\zeta}-\Omega_{\phi\zeta}(\rho_{\phi\phi}-\rho_{\zeta\zeta}) +\Omega_{\phi\xi}\rho_{\xi\zeta} -\Omega_{\psi\zeta}\rho_{\phi\psi} ,\nonumber\\
\dot{\rho}_{\xi\zeta} =& -\left[\frac{1}{2}\Gamma_{\chi}(1+2u^{2}) +2i(\alpha^{2}-\beta^{2})\Delta\right]\rho_{\xi\zeta}\nonumber\\
&+\frac{1}{2}\left(\sqrt{\Gamma_{\xi\phi}\Gamma_{\zeta\phi}} +\sqrt{\Gamma_{\xi\psi}\Gamma_{\zeta\psi}}\right) (\rho_{\zeta\zeta} +\rho_{\xi\xi}) \nonumber\\
&+ \left(\sqrt{2}G_{\chi}-2i\alpha\beta\Delta\right) (\rho_{\zeta\zeta} - \rho_{\xi\xi}) \nonumber\\
&-\Omega_{\phi\xi}\rho_{\phi\zeta} -\Omega_{\psi\xi}\rho_{\psi\zeta} -\Omega_{\psi\zeta}\rho_{\xi\psi} -\Omega_{\phi\zeta}\rho_{\xi\phi} \nonumber\\
&-(\alpha \tilde{E}\rho_{1\zeta}+\beta \tilde{E}^{\ast}\rho_{\xi1})  , \label{C3}
\end{align}
where $\tilde{E}=E/\sqrt{2}$ and
\begin{align}
\Omega_{\psi\xi} &= \left(\sqrt{2}u\alpha +w\beta\right)\Omega_{c} + u\beta\Omega_{a} ,\nonumber\\
\Omega_{\psi\zeta} &= \left(\sqrt{2}u\beta -w\alpha\right)\Omega_{c} - u\alpha\Omega_{a} ,\nonumber\\
\Omega_{\phi\xi} &= \left(\sqrt{2}w\alpha -u\beta\right)\Omega_{c} + w\beta\Omega_{a} ,\nonumber\\
\Omega_{\phi\zeta} &= \left(\sqrt{2}w\beta +u\alpha\right)\Omega_{c} - w\alpha\Omega_{a} ,\label{C4}
\end{align}
are the effective Rabi frequencies of the driving fields between the single and double excitation states.


\begin{thebibliography}{21}

\bibitem{lm16} P. Lodahl, S. Mahmoodian, S. Stobbe, P. Schneeweiss, J. Volz, A. Rauschenbeutel, H. Pichler, and P. Zoller, Nature (London) {\bf 541}, 473 (2017), and references therein.

\bibitem{vr16} B. Vermersch, T. Ramos, P. Hauke, and P. Zoller, Phys. Rev. A {\bf 93}, 063830 (2016).

\bibitem{pv14} J. Petersen, J. Volz, and A. Rauschenbeutel, Science {\bf 346}, 67 (2014).

\bibitem{ms14} R. Mitsch, C. Sayrin, B. Albrecht, P. Schneeweiss, and A. Rauschenbeutel, Nat. Commun. {\bf 5}, 5713 (2014).

\bibitem{sn15} I. S\"{o}llner, S. Mahmoodian, S. L. Hansen, L. Midolo, A. Javadi, G. Kirsanske, T. Pregnolato, H. El-Ella, E. H. Lee, J. D. Song, S. Stobbe, and P. Lodahl, Nature Nanotech. {\bf 10}, 775 (2015).

\bibitem{yt15} A. B. Young, A. C. T. Thijssen, D. M. Beggs, P. Androvitsaneas, L. Kuipers, J. G. Rarity, S. Hughes, and R. Oulton, Phys. Rev. Lett. {\bf 115}, 153901 (2015).

\bibitem{nb14} M. Neugebauer, T. Bauer, P. Banzer, and G. Leuchs, Nano Lett. {\bf 14}, 2546 (2014).

\bibitem{rp14} T. Ramos, H. Pichler, A. J. Daley, and P.  Zoller, Phys. Rev. Lett. {\bf 113}, 237203 (2014).

\bibitem{pr15} H. Pichler, T. Ramos, A. J. Daley, and P. Zoller, Phys. Rev. A {\bf 91}, 042116 (2015).


%\bibitem{ki15}S. Schneider and G. J. Millburn, Phys. Rev. A {\bf 65}, 042107 (2002).

\bibitem{sr12} K. Stannigel, P. Rabl, and P. Zoller, New J. Phys. {\bf 14}, 063014 (2012).

\bibitem{ki14}B. Kraus, H. P. B\"{u}chler, S. Diehl, A. Kantian, A. Micheli, and P. Zoller, Phys. Rev. A {\bf 78}, 042307 (2008).

\bibitem{hk08} H. J. Kimble, Nature (London) {\bf 453}, 1023 (2008).

\bibitem{wg93} C. W. Gardiner, Phys. Rev. Lett. {\bf 70}, 2269 (1993).

\bibitem{hc93} H. J. Carmichael, Phys. Rev. Lett. {\bf 70}, 2273 (1993).

\bibitem{kc94} P. Kochan and H. J. Carmichael, Phys. Rev. A {\bf 50}, 1700 (1994).

\bibitem{gp94} C. W. Gardiner and A. S. Parkins, Phys. Rev. A {\bf 50}, 1792 (1994).

\bibitem{nc05} H. Nha and H. J. Carmichael, Phys. Rev. A {\bf 71}, 013805 (2005).

\bibitem{gz04} C.W. Gardiner and P. Zoller, {\it Quantum Noise}, 3rd ed. (Springer-Verlag, Berlin, 2004), chap. 12.

\bibitem{ki11} C. Fabre, M. Pinard, S. Bourzeix, A. Heidmann, E. Giacobino, and S. Reynaud, Phys. Rev. A {\bf 49}, 1337 (1994).

\bibitem{ki12}I. Wilson-Rae, P. Zoller, and A. Imamo\={g}lu, Phys. Rev. Lett. {\bf 92}, 075507 (2004).

\bibitem{ki13} J. M\'{a}rquez, L. Geelhaar, and K. Jacobi, Appl. Phys. Lett. {\bf 78}, 2309 (2001).

\bibitem{tl94} W. Leo\'nski and R. Tana\'s, Phys. Rev. A {\bf 49}, R20 (1994).

\bibitem{is97} A. Imamoglu, H. Schmidt, G. Woods, and M. Deutsch, Phys. Rev. Lett. {\bf 79}, 1467 (1997).

\bibitem{hs11} A. J. Hoffman, S. J. Srinivasan, S. Schmidt, L. Spietz, J. Aumentado, H. E. T\"{u}reci, and A. A. Houck, Phys. Rev. Lett. {\bf 107}, 053602 (2011).

\bibitem{bi11} M. Bamba, A. Imamoglu, I. Carusotto, and C. Ciuti, Phys. Rev. A {\bf 83}, 021802(R) (2011).

\bibitem{kk16} J. K. Kalaga, A. Kowalewska-Kudlaszyk, W. Leo\'nski, and A. Barasi\'nski, Phys. Rev. A {\bf 94}, 032304 (2016).

\bibitem{ki01} J. H. Li, R. Yu,  and Y. Wu, Phys. Rev. A {\bf 94}, 053837 (2015).

\bibitem{ki03} B. Y. Zhou and G. X. Li, Phys. Rev. A {\bf 94}, 033809 (2016).

\bibitem{ki05} W. W. Deng, G. X. Li, and H. Qin, Phys. Rev. A {\bf 91}, 043831 (2015).

\bibitem{ki06} G. Q. Yang, W. J. Gu, G. X. Li, B. C. Zhou, and Y. F. Zhu, Phys. Rev. A {\bf 92}, 033822 (2015).

\bibitem{tc92} L. Tian and H. J. Carmichael, Phys. Rev. A {\bf 46}, R6801 (1992).

\bibitem{bb05} K. M. Birnbaum, A. Boca, R. Miller, A. D. Boozer, T. E. Northup, and H. J. Kimble, Nature (London) {\bf 436}, 87 (2005).


\bibitem{cp10} D. Comparat and P. Pillet, J. Opt. Soc. Am. B {\bf 27}, A208 (2010), and references therein.

\bibitem{lf01} M. D. Lukin, M. Fleischhauer, R. Cote, L. M. Duan, D. Jaksch, J. I. Cirac, and P. Zoller, Phys. Rev. Lett. \textbf{87}, 037901 (2001).

\bibitem{uj09} E. Urban, T. A. Johnson, T. Henage, L. Isenhower, D. D. Yavuz, T. G. Walker, and M. Saffman, Nat. Phys. \textbf{5}, 110 (2009).

\bibitem{ga10} J. Gillet, G. S. Agarwal, and T. Bastin, Phys. Rev. A \textbf{81}, 013837 (2010).

\bibitem{ge10} A. Ga\"etan, C. Evellin, J. Wolters, P. Grangier, T. Wilk, and A. Browaeys, New J. Phys. {\bf 12}, 065040 (2010).

\bibitem{hh10} Y. Han, B. He, K. Heshami, C.-Z. Li, and C. Simon, Phys. Rev. A {\bf 81}, 052311 (2010).

\bibitem{zm10} B. Zhao, M. M\"uller, K. Hammerer, and P. Zoller, Phys. Rev. A {\bf 81}, 052329 (2010).

\bibitem{nm10} A. E. B. Nielsen and K. Molmer, Phys. Rev. A {\bf 82}, 052326 (2010).

\bibitem{ss11} L. Salen, S. I. Simonsen, and J. P. Hansen, Phys. Rev. A \textbf{83}, 015401 (2011).

\bibitem{at11} K. Almutairi, R. Tana\'s, and Z. Ficek, Phys. Rev. A {\bf 84}, 013831 (2011).

\bibitem{jr10} J. E. Johnson and S. L. Rolston, Phys. Rev. A {\bf 82}, 033412 (2010).

\bibitem{gm09} A. Ga\"etan, Y. Miroshnychenko, T. Wilk, A. Chotia, M. Viteau, D. Comarat, P. Pillet, A. Browaeys, and P. Grangier, Nat. Phys. \textbf{5}, 115 (2009).

\bibitem{hr07} R. Heidemann, U. Raitzsch, V. Bendkowsky, B. Butscher, R. L\"ow, L. Santos, and T. Pfau, Phys. Rev. Lett. \textbf{99}, 163601 (2007).

\bibitem{al86} D. V. Averin and K. K. Likharev, J. Low Temp. Phys. {\bf 62}, 345 (1986).

\bibitem{fd87} T. A. Fulton and G. J. Dolan, Phys. Rev. Lett. {\bf 59}, 109 (1987).

\bibitem{mk92} M. Kastner, Rev. Mod. Phys. {\bf 64}, 849 (1992).

\bibitem{ps96} G. M. Palma, K.-A. Suominen, and A. K. Ekert, Proc. R. Soc. London, Ser. A {\bf 452}, 567 (1996).

\bibitem{dg97} L.-M. Duan and G.-C. Guo, Phys. Rev. Lett. {\bf 79}, 1953 (1997).

\bibitem{zr97} P. Zanardi and M. Rasetti, Phys. Rev. Lett. {\bf 79}, 3306 (1997).

\bibitem{lc98} D. A. Lidar, I. L. Chuang, and K. B. Whaley, Phys. Rev. Lett. {\bf 81}, 2594 (1998).

\bibitem{kl00} E. Knill, R. Laflamme, and L. Viola, Phys. Rev. Lett. {\bf 84}, 2525 (2000).

\bibitem{pb07} P. G. Brooke, Phys. Rev. A {\bf 75}, 022320 (2007).

\bibitem{wk86} L.-A. Wu, H. J. Kimble, J. L. Hall, and H. Wu, Phys. Rev. Lett. {\bf 57}, 2520 (1986).

\bibitem{cj12} D. E. Chang, L. Jiang, A. V. Gorshkov, and H. J. Kimble, New J. Phys. {\bf 14}, 063003 (2012).

\bibitem{gp13} A. Gonzalez-Tudela and D. Porras, Phys. Rev. Lett. {\bf 110}, 080502 (2013).

\bibitem{ll07} G. Lecamp, P. Lalanne, and J. P. Hugonin, Phys. Rev. Lett. {\bf 99}, 023902 (2007).

\bibitem{jo13} C. Junge, D. O'Shea, J. Volz, and A. Rauschenbeutel, Phys. Rev. Lett. {\bf 110}, 213604 (2013).

\bibitem{as14} M. Arcari, {\it et al.}, Phys. Rev. Lett. {\bf 113}, 093603 (2014).

\bibitem{bb00} A. Beige, D. Braun, and P. L. Knight, New J. Phys. {\bf 2}, 22 (2000).

\bibitem{lw03} D. A. Lidar and K. B. Whaley, in {\it Irreversible Quantum Dynamics}, edited by F. Benatti and R. Floreanini (Springer, Berlin, 2003), p. 83.

\bibitem{cw76} H. J. Carmichael and D. F. Walls, J. Phys. B {\bf 9}, 1199 (1976).

\bibitem{km76} H. J. Kimble and L. Mandel, Phys. Rev. A {\bf 13}, 2123 (1976).

\bibitem{kd77} H. J. Kimble, M. Dagenais, and L. Mandel, Phys. Rev. Lett. {\bf 39}, 691 (1977).

\bibitem{ta99} S. M. Tan, J. Opt. B: Quantum Semiclass. Opt. {\bf 1}, 424 (1999).

\bibitem{ks14} O. Kyriienko, I. A. Shelykh, and T. C. H. Liew, Phys. Rev. A {\bf 90}, 033807 (2014).

\bibitem{fs16} H. Flayac and V. Savona, Phys. Rev. A {\bf 94}, 013815 (2016).

\bibitem{lh70} R. H. Lehmberg, Phys. Rev. A {\bf 2}, 883 (1970).

\bibitem{lo73} W. H. Louisell, {\it Quantum Statistical Properties of Radiation}, (Wiley, New York, 1973), chap. 6.


\end{thebibliography}
\end{document}